\def\mycmd{2}

\if\mycmd1
\documentclass[11pt,onecolumn, draftcls]{IEEEtran}
\else 
\documentclass[journal]{IEEEtran}
\fi
\usepackage[utf8]{inputenc}
\usepackage{colortbl}    
\usepackage{pifont}     

\usepackage[british]{babel}
\usepackage{multicol}
\usepackage{array,ragged2e}
\usepackage{float}
\usepackage{mathtools}
\usepackage{amsmath}
\usepackage{amsthm}
\usepackage{amssymb}
\usepackage{graphicx}
\usepackage{setspace}
\usepackage{color}
\usepackage{bm}
\usepackage{kotex}
\usepackage{cases}
\usepackage{makecell}
\usepackage{enumitem}
\usepackage{flowchart}
\usepackage{amsfonts}
\usepackage{steinmetz}
\usetikzlibrary{matrix,shapes,arrows,positioning,chains}
\usepackage{lmodern,babel,adjustbox,booktabs,multirow}
\usepackage{makecell}
\usepackage{pbox}
\usepackage{epsfig}
\usepackage{subfigure}

\makeatletter
\newcommand{\multiline}[1]{%
  \begin{tabularx}{\dimexpr\linewidth-\ALG@thistlm}[t]{@{}X@{}}
    #1
  \end{tabularx}
}

\floatstyle{ruled}
\newfloat{algorithm}{tbp}{loa}
\providecommand{\algorithmname}{Algorithm}
\floatname{algorithm}{\protect\algorithmname}

\theoremstyle{plain}

\theoremstyle{plain}

\theoremstyle{plain}

\makeatletter
\renewcommand\paragraph[1]{%
    \vspace{.1cm}\noindent\textbf{#1.}
}
\makeatother

\usepackage{epsfig}
\usepackage[caption=false,font=normalsize,labelfont=sf,textfont=sf]{subfig}
\usepackage{cite}
\usepackage{stfloats}
\usepackage{graphicx}
\usepackage{multirow}
\usepackage{array}
\usepackage{enumerate}
\usepackage{graphicx}
\usepackage{times}
\usepackage{algorithm}
\usepackage{algpseudocode}
\theoremstyle{remark}

\newtheorem{lemma}{Lemma}

\newtheorem{proposition}{Proposition}

\makeatother

\algrenewcommand\algorithmicindent{1.0em}%
\providecommand{\lemmaname}{Lemma}
\providecommand{\propositionname}{Proposition}

\providecommand{\theoremname}{Theorem}
\providecommand{\theoremname}{Definition}
\newcommand{\rom}[1]{\uppercase\expandafter{\romannumeral #1\relax}}

\newcounter{problem}
\newcounter{save@equation}
\newcounter{save@problem}

\definecolor{lightergray}{gray}{0.8}
\definecolor{ForestGreen}{RGB}{34,139,34}  

\newcommand{\Xmark}{\textcolor{lightergray}{\ding{55}}}
\newcommand{\mycheck}{\textcolor{ForestGreen}{\ding{51}}}

\newcommand{\Problem}[1]{\bf{\mathdutchcal{P}#1}}
\DeclareMathAlphabet{\mathdutchcal}{U}{eus}{b}{n}
\DeclareMathAlphabet{\mathcal}{OMS}{cmsy}{m}{n}
\SetMathAlphabet{\mathcal}{bold}{OMS}{cmsy}{b}{n}


\numberwithin{save@problem}{subsection}
\numberwithin{save@equation}{subsection}

\allowdisplaybreaks[4]

\begin{document}
\title{Multiple Active STAR-RIS-Assisted Secure Integrated Sensing and Communication via Cooperative Beamforming}
\author{Hyeonho Noh,~\IEEEmembership{Member,~IEEE}, Hyeonsu Lyu,~\IEEEmembership{Student Member,~IEEE}, and Hyun Jong Yang,~\IEEEmembership{Senior Member,~IEEE}
\thanks{Hyeonho Noh is with Electronics and Telecommunications Research Institute (ETRI), Korea (e-mail: hhnoh@etri.re.kr).
Hyeonsu Lyu is with the Department of Electrical Engineering, POSTECH, Korea (e-mail: hslyu4@postech.ac.kr). 
Hyun Jong Yang is with the Department of Electrical and Computer Engineering, Seoul National University, Korea (email: hjyang@snu.ac.kr).
}
}

\maketitle
\begin{abstract}\label{abstract}
This paper explores an integrated sensing and communication (ISAC) network empowered by multiple active simultaneously transmitting and reflecting reconfigurable intelligent surfaces (STAR-RISs). A base station (BS) furnishes downlink communication to multiple users while concurrently interrogating a sensing target. We jointly optimize the BS transmit beamformer and the reflection/transmission coefficients of every active STAR-RIS in order to maximize the aggregate communication sum-rate, subject to (i) a stringent sensing signal-to-interference-plus-noise ratio (SINR) requirement, (ii) an upper bound on the leakage of confidential information, and (iii) individual hardware and total power constraints at both the BS and the STAR-RISs.
The resulting highly non-convex program is tackled with an efficient alternating optimization (AO) framework. First, the original formulation is reformulated into an equivalent yet more tractable representation and partitioned into sub-problems. The BS beamformer is updated in closed form via the Karush–Kuhn–Tucker (KKT) conditions, whereas the STAR-RIS reflection and transmission vectors are refined through successive convex approximation (SCA), yielding a semidefinite program that is then solved via semidefinite relaxation.
Comprehensive simulations demonstrate that the proposed algorithm delivers substantial sum-rate gains over passive-RIS and single STAR-RIS baselines, all the while rigorously meeting the prescribed sensing and security constraints.



\end{abstract}


\section{Introduction}
\label{sec:introduction}

Integrated sensing and communication (ISAC) is regarded as one of the crucial technologies for 6G. 
ISAC integrates radar sensing and wireless communication on a single hardware platform, using the same radio and time resources to perform both functions simultaneously.
This dual-function radar-communication approach not only reduces hardware costs and form factor but also minimizes resource usage compared to the separate deployment of individual radar and communication systems. As a result, ISAC underpins numerous future applications, including in-home sensing, smart-city services, and digital twin technologies.

However, current ISAC systems face several notable challenges, including blind spots arising from line-of-sight (LoS) constraints in sensing, mutual interference between sensing and communication due to shared spectrum, performance trade-offs, and potential security issues associated with communication symbol-embedded waveforms. To address these limitations, reconfigurable intelligent surfaces (RISs) have attracted significant research interest. By adjusting the wireless propagation environment to create favorable LoS links and provide additional degrees of freedom (DoF), RISs can mitigate LoS blockages and form an enhanced radio environment. As metasurfaces capable of actively controlling the phase and amplitude of electromagnetic waves, RISs go far beyond conventional passive reflectors, enhancing both sensing and communication performance.

In recent years, substantial efforts have been made to integrate RIS into ISAC systems by jointly optimizing the transmit beamforming at the base station (BS) and the phase shifts of the RIS \cite{Xinyi21_WCL, Xinyi22_TVT, Ziyi22_JSAC, Yuan24_TWC, Xudong24_IOTJ, Yongqing25_JSAC, Lyu24_TVT, Guang24_TVT}. 
One of the earliest works, \cite{Xinyi21_WCL}, introduces a joint design framework to maximize radar detection probability under power and quality-of-service (QoS) constraints, using semidefinite programming (SDP) and relaxation techniques. This is extended in \cite{Xinyi22_TVT} to account for constant-modulus waveforms and discrete RIS phase shifts for multi-target direction-of-arrival estimation. In \cite{Yuan24_TWC}, a full-duplex ISAC system is proposed, where uplink communication and target sensing are performed simultaneously via joint optimization of beamforming, RIS control, and user power allocation. A deep reinforcement learning (DRL)-based solution is presented in \cite{Xudong24_IOTJ} for real-time optimization in RIS-assisted ISAC scenarios. Further studies explore cognitive radio integration \cite{Yongqing25_JSAC}, hybrid orthogonal/non-orthogonal multiple access schemes \cite{Lyu24_TVT}, and advanced beyond-diagonal RIS architectures \cite{Guang24_TVT}, highlighting the versatility of RIS in future wireless sensing-communication networks.

Despite the rich body of literature on applying RISs to ISAC, there are several key challenges:
\begin{itemize}[leftmargin=*]
    \item \textbf{Coverage}: Most RIS-ISAC studies assume that the users and sensing targets lie in the same half-space defined by the BS and RIS, which limits the overall ISAC service coverage.
    \item \textbf{Multi-hop fading}: For NLoS targets, sensing signals can experience multiple RIS reflections, causing severe signal-to-noise ratio (SNR) loss from multi-hop fading. Conventional beamforming and RIS phase optimization cannot effectively address the problem.
    \item \textbf{Security}: ISAC waveforms inherently embed communication symbols in sensing directions. Consequently, if an ISAC waveform is directed or leaked toward an eavesdropper, serious privacy and security issues may arise.
\end{itemize}

\subsection{Related Works}

\paragraph{Simultaneously transmitting and reflecting RIS (STAR-RIS)}
STAR-RISs have been proposed \cite{Liu21_WC} as a solution to the half-space coverage limitation of conventional RISs. By splitting an incident signal into reflected and transmitted components, a STAR-RIS achieves full-space coverage.
In \cite{Zhipeng23_TVT}, a joint optimization framework is proposed to maximize sensing signal-to-noise ratio while ensuring communication QoS by jointly designing BS beamforming and STAR-RIS transmission and reflection (T\&R) coefficients.
The work in \cite{Qin25_CL} further incorporates deployment location optimization under an energy efficiency objective. 
The study in \cite{Xue24_TWC} explores STAR-RIS ISAC systems with non-orthogonal multiple access (NOMA) to enhance beampattern design under communication and power constraints. In addition, low-cost sensing hardware \cite{Wang23_TWC} and secure transmission strategies \cite{Wei24_TWC} have been proposed to address practical challenges in separating sensing and communication regions. To reduce hardware cost, \cite{Zhang24_TWC_decoupling} considers coupled phase constraints and proposes alternating direction method of multipliers- and DRL-based algorithms for joint waveform and STAR-RIS configuration.

\paragraph{Active RIS}
Active RIS, which can amplify signals rather than merely reflect them, has also attracted attention for ISAC. 
\cite{Zhu23_TVT} and \cite{Yu24_TCOM} introduce active RISs to maximize the radar output SNR while guaranteeing predefined signal-to-interference-plus-noise ratio (SINR) for communication users. 
Similarly, \cite{Zhu24_TWC} demonstrates that active RISs enhance target angle estimation and ensure robust communication performance.
\cite{Zhang24_TWC} introduces active STAR-RIS into ISAC systems for the first time, deriving optimal active STAR-RIS coefficients under three different hardware constraints. 
Furthermore, \cite{Dongsheng24_TWC} embeds low-cost sensing elements into active STAR-RIS to directly receive target echoes—mitigating multi-hop attenuation—and jointly optimizes the sensing filter design, transmit beamforming, and active RIS coefficients to maximize echo SINR.
In addition, \cite{Hao23_WCL} and \cite{Sravani24_TWC} investigate THz ISAC systems with dynamic delay alignment modulation based on active RISs, effectively synchronizing multi-path signals and handling mobility challenges.

\paragraph{Cooperative RIS}
Cooperative deployments of multiple RISs have been investigated to further enhance ISAC performance.
\cite{Yinghui22_JSAC} proposes a double-RIS ISAC system: a transmitter-side RIS suppresses interference toward the radar, while a receiver-side RIS cancels radar-induced interference, jointly maximizing communication SINR while preserving radar performance.
\cite{Yu22_TSP} proposes a distributed semi-passive RIS framework that splits operation into an ISAC phase for user localization and a communication phase, each supported by a distinct RIS.
In \cite{Chen24_TGCN}, two cooperative RISs jointly optimize passive beamforming to maximize energy efficiency under user data-rate constraints. Expanding beyond dual-RIS scenarios, \cite{Xiaoyan22_TWC} considers systems with multiple RISs and proposes joint optimization of transmit beamforming and RIS phase shifts. Multi-hop user links via RISs are explored in \cite{Ma24_TWC} using graph-theoretic methods, while \cite{Shaikh24_TVT} and \cite{Saqib24_TWC} introduce broader frameworks that incorporate user clustering, STAR-RIS association, power control, and RIS placement.

\paragraph{Secure ISAC}
Security has emerged as a critical concern in ISAC systems, where the sensing target may wiretap the communication signals. In \cite{Chengjun24_IOTJ}, a secure beamforming scheme is proposed for a RIS-assisted NOMA-ISAC system, jointly optimizing BS and RIS beamforming to maximize the sum secrecy rate under QoS, power, and sensing constraints. Robust secure ISAC with channel uncertainty is addressed in \cite{Tao24_IoTJ}, where BS beamforming, STAR-RIS coefficients, and artificial noise are optimized. Active RISs are leveraged in \cite{Chen23_GLOBECOM} to enable covert communications in NOMA-ISAC, outperforming passive RIS in secrecy and sensing performance. \cite{Yang24_TVT} introduces an interference cancellation strategy using active RISs to mitigate eavesdropping while maintaining sensing capability. In \cite{Zhao23_CL}, high-power radar signals reflected via RIS disrupt eavesdropper reception without dedicated jamming. Lastly, \cite{Liu25_IoTJ} proposes an ISAC framework that incorporates detection error probability constraints to protect against both eavesdropping and sensing-based attacks.

\begin{table*}
\centering
    \caption{Comparison of the proposed scheme to recent works.}
    \adjustbox{width= \if 1\mycmd 0.6 \else 1.0 \fi \textwidth}{
    \begin{tabular}{cccccccc}
    \toprule
    \textbf{Ref.} & \textbf{RIS type} & \textbf{Power type} & \textbf{Multiple RIS} & \textbf{Security} & \textbf{Comm. metric} & \textbf{Sensing metric} & \textbf{Target location} \\ \midrule[\heavyrulewidth]
    \arrayrulecolor{lightergray}
    \cite{Xudong24_IOTJ, Yongqing25_JSAC} & RIS & \textcolor{lightergray}{Passive} & \Xmark & \Xmark & QoS & PEB & NLoS \\ \cline{1-8}
    \cite{Qin25_CL} & STAR-RIS & \textcolor{lightergray}{Passive} & \Xmark & \mycheck & SINR & Beampattern gain & NLoS \\ \cline{1-8}
    \cite{Zhipeng23_TVT} & STAR-RIS & \textcolor{lightergray}{Passive} & \Xmark & \Xmark & SINR & Echo SNR & NLoS \\ \cline{1-8}
    \cite{Yufei24_TWC} & STAR-RIS & \textcolor{lightergray}{Passive} & \Xmark & \Xmark & SINR & Echo SINR & NLoS \\ \cline{1-8}
    \cite{Yinghui22_JSAC} & RIS & \textcolor{lightergray}{Passive} & \mycheck & \Xmark & SINR & Echo SINR & NLoS \\ \cline{1-8}
    \cite{Tao24_IoTJ} & STAR-RIS & \textcolor{lightergray}{Passive} & \Xmark & \Xmark & Sum-rate & Waveform & NLoS \\ \cline{1-8}
    \cite{Yuan24_TWC} & RIS & \textcolor{lightergray}{Passive} & \Xmark & \Xmark & Sum-rate & Echo SINR & LoS \\ \cline{1-8}
    \cite{Chengjun24_IOTJ} & RIS & \textcolor{lightergray}{Passive} & \Xmark & \mycheck & Sum-rate & Beampattern gain & LoS \\ \cline{1-8}
    \cite{Qing24_TGCN, Xiaoyan22_TWC, Xiaoyan24_TWC} & RIS & \textcolor{lightergray}{Passive} & \mycheck & \Xmark & Sum-rate & - & - \\ \cline{1-8}
    \cite{Naim24_TVT} & STAR-RIS & \textcolor{lightergray}{Passive} & \mycheck & \Xmark & Sum-rate & - & - \\ \cline{1-8}
    \cite{Sravani24_TWC} & STAR-RIS & Active & \Xmark & \Xmark & Sum-rate & SNR & LoS \\ \cline{1-8}
    \cite{Dongsheng24_TWC} & STAR-RIS & Active & \Xmark & \Xmark & QoS & Echo SINR & NLoS \\ \cline{1-8}
    \cite{Hao24_TVT} & RIS & Active & \Xmark & \mycheck & QoS & Echo SINR & NLoS \\
    \arrayrulecolor{black}
    \midrule[\heavyrulewidth]
    \textbf{Ours} & \textbf{STAR-RIS} & \textbf{Active} & \textbf{\mycheck} & \textbf{\mycheck} & \textbf{Sum-rate} & \textbf{Echo SINR} & \textbf{NLoS} \\ 
    \bottomrule
    \end{tabular}
    }
    \label{tab:compare}
\end{table*}

\subsection{Main Contributions} 
\label{subsec:contributions}
Despite the aforementioned approaches, a comprehensive solution that simultaneously addresses the challenges of ISAC systems remains largely unexplored.
This calls for deeper exploration of next-generation RIS architectures with advanced joint optimization systems.
Therefore, this paper introduces a secure, cooperative, and active STAR-RIS ISAC framework. The proposed system deploys multiple active STAR-RISs to simultaneously support multi-user communication and enable target sensing. 
The contributions are summarized as follows:

\begin{itemize}[leftmargin=*]
\item \textbf{Full-Featured RIS-ISAC Framework:}
We design a multi-STAR-RIS ISAC framework to create full-space LoS coverage for communication users and the sensing target, which is \textit{\textbf{the first attempt to integrate  active and cooperative STAR-RISs into a secure ISAC system.}}
The active and cooperative STAR-RIS relays provide abundant link diversity and reliability to communication users and sensing target while the active gain can compensate for the multi-hop path loss. At the same time, as the sensing target can potentially act as an eavesdropper, we secure the ISAC system by deploying a dedicated sensing signal and beamforming vector, which increases the target signal's SINR while suppressing the communication signals' SINR at the sensing target. 

\item \textbf{Proposition of Open STAR-RIS-ISAC Problem:} We formulate a sum-rate maximization problem that jointly optimizes the transmit beamforming of the BS and T\&R coefficients of the passive and active STAR-RISs.
This optimization is characterized by the three constraints: i) the minimum SINR of the echo sensing signal; ii) the maximum tolerable information leakage toward the sensing target; and iii) the total power constraint of STAR-RISs.

~~~Then, the problem is intractable due to the non-convex objective and tight couplings of beamforming vectors in the SINR.
Thus, we reformulate the problem into a tractable convex problem by introducing slack variables. 
We then decompose the overall problem into three sub-problems: the sensing beamformer, communication beamformer, and the T\&R beamformer.
Through the proposed alternating optimization (AO) approach, we design the low-complexity algorithm with robust ISAC performance.

\item \textbf{Low-Complexity Beamforming Design:} 
We first derive a power-constrained minimum variance distortionless response (MVDR) beamformer for the sensing target. By introducing a Lagrange multiplier to handle the power constraint, we find the optimal solution with linear computational complexity. 
Next, we utilize the Karush–Kuhn–Tucker (KKT) conditions and block coordinate descent to obtain the communication beamformer. 
Finally, to optimize the T\&R beamformer at each STAR-RIS, we fix the beamformers of all other STAR-RISs and sequentially update the specific target STAR-RIS.
Each user’s channel is therefore reformulated so that it depends only on the amplitude and phase-shift parameters of that single STAR-RIS.
Then, the resulting single STAR-RIS problem is converted into a convex problem via successive convex approximation (SCA) and semidefinite relaxation (SDR), which can be easily solved via convex programming.
\item \textbf{Observation of a Secure, Cooperative, and Active STAR-RIS}:
Numerical simulations confirm that the proposed cooperative and active STAR-RIS achieves superior performance compared to a single or passive STAR-RIS, showing that active STAR-RIS can match passive STAR-RIS performance with fewer RIS units.
Hence, cooperative and active STAR-RIS is verified as an effective means to address multi-hop fading in ISAC systems.
Moreover, we verify that the system can secure the communication against the sensing target, enhancing the overall security performance.
\end{itemize}

\textit{Notations}: $(\cdot)^{*}$, $(\cdot)^\mathsf{T}$, and $(\cdot)^\mathsf{H}$ are complex
conjugate, transpose, and conjugate transpose, respectively. $\mathbf{A}^{-1}$ and $|\mathbf{A}|$ are the inverse and absolute of matrix $\mathbf{A}$, respectively. $\mathbf{I}_N$ is the $N \times N$ identity matrix. The diagonal matrix with diagonal elements $(a_1, a_2, \ldots,a_m)$ is denoted by $\mathrm{diag}(a_1, a_2, \ldots, a_m)$. 
The complex normal distribution is denoted by $\mathcal{CN} (\mu, \Sigma)$ with mean $\mu$ and covariance matrix $\Sigma$. $\mathbb{R}^{n \times m}$ and $\mathbb{C}^{n \times m}$ are the $(n \times m)$-dimensional real and complex spaces, respectively. $\mathbb{E}(\cdot)$ is the expectation.  $\left\| \cdot \right\|_\text{F}$ is the Frobenius norm.

\section{System Model}
\label{sec:system_model}

\begin{figure}[t!]
    \centering
    \includegraphics[draft=false, width= \if 1\mycmd 1.0 \else 0.95 \fi \columnwidth]{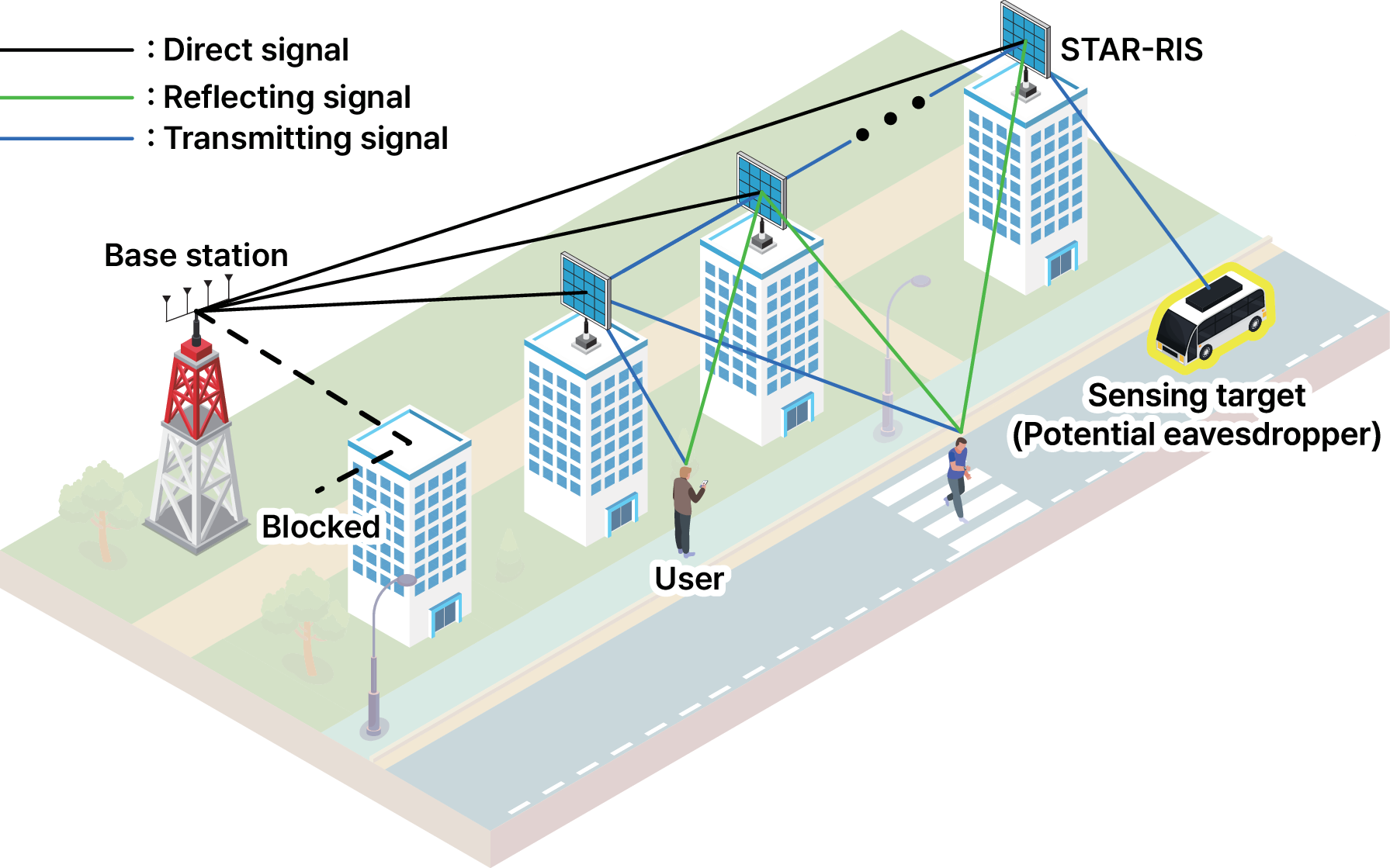}
    \vspace{-5pt}
    \caption{Illustration of the proposed multiple STAR-RIS scenario.}
    \label{fig:system_model}
    \vspace{-10pt}
\end{figure}

Fig.~\ref{fig:system_model} illustrates the proposed ISAC system, where a single ISAC BS equipped with a uniform linear array consisting of $N$ transmit antennas transmits ISAC signals to $K$ single-antenna communication users for downlink communication and to a region of interest for localizing a sensing target.
We assume that the direct link between the BS and communication users (or sensing target) is blocked, i.e., an NLoS environment.
To improve system performance by establishing LoS links between the BS and users (or sensing targets), we deploy $L$ distributed STAR-RISs, each equipped with a uniform planar array (UPA) of $M$ T\&R elements.
Without loss of generality, we index the STAR-RISs from $1$ to $L$ in ascending order of proximity to the BS, assigning lower indices to those closer to the BS.
The incident ISAC signal from the BS reaches STAR-RIS~$1$, and then splits into T\&R signals.  
The transmission signal is then forwarded to the next STAR-RIS (e.g., STAR-RIS~$2$) for further communication and sensing, while the reflection signal is directed to the communication users.
After the signal reaches the $L$-th STAR-RIS, the transmission signal arrives at the sensing target in the region of interest. 
The signal reflected by the target then propagates back through the STAR-RISs in reverse order, and is finally received by the BS for sensing.

\subsection{STAR-RIS, Signal, and Channel Models}
Let $\boldsymbol{\Theta}_{\text{T},l}, \boldsymbol{\Theta}_{\text{R},l} \in \mathbb{C}^{M \times M}$ denote the T\&R coefficient matrices of the $l$-th STAR-RIS, respectively. Each matrix can be modeled as $\boldsymbol{\Theta}_{m,l} = \mathbf{B}_{m,l} \boldsymbol{\Phi}_{m,l}$ for $m \in \{ \text{T}, \text{R} \}$, where $\mathbf{B}_{m,l} = \mathrm{diag}(\boldsymbol{\beta}_{m,l})$ is the amplitude matrix with $\boldsymbol{\beta}_{m,l} = [\beta_{m,l,1}, \ldots, \beta_{m,l,M}]^\mathsf{T}$ and $\boldsymbol{\Phi}_{m,l} = \mathrm{diag}(\boldsymbol{\phi}_{m,l})$ is the phase shift matrix at the $l$-th STAR-RIS with $\boldsymbol{\phi}_{m,l} = [e^{j \phi_{m,l,1}}, \ldots, e^{j \phi_{m,l,M}}]^\mathsf{T}$ for $m \in \{ \text{T}, \text{R} \}$.

The complex-valued transmit signal is written by
\begin{align}
\mathbf{x} = \sum_{k \in \mathcal{K}} \mathbf{w}_{k} c_{k} + \mathbf{w}_S s, 
\end{align}
where $\mathbf{w}_{k}$ and $\mathbf{w}_S$ are the transmit beamforming vector for the $k$-th communication user and sensing target, respectively. $c_{k}$ is the symbol for the $k$-th communication user, and $s$ is the radar probing symbol. 
It is assumed that communication symbols $c_k$ are independent and identically distributed with zero mean and unit variance for all $k \in \mathcal{K}$, the radar probing symbol is generated by pseudo-random coding, i.e., $\mathbb{E} [c_i^* c_j] = 0$ for $i \neq j$ and $\mathbb{E}[c_k^* s] = 0$, and $\mathcal{K} = \{ 1, \ldots, K \}$ with $S \notin \mathcal{K}$.

\subsection{Communication and Sensing Performance}

\begin{figure}[t!]
    \centering
    \includegraphics[width=0.95\linewidth]{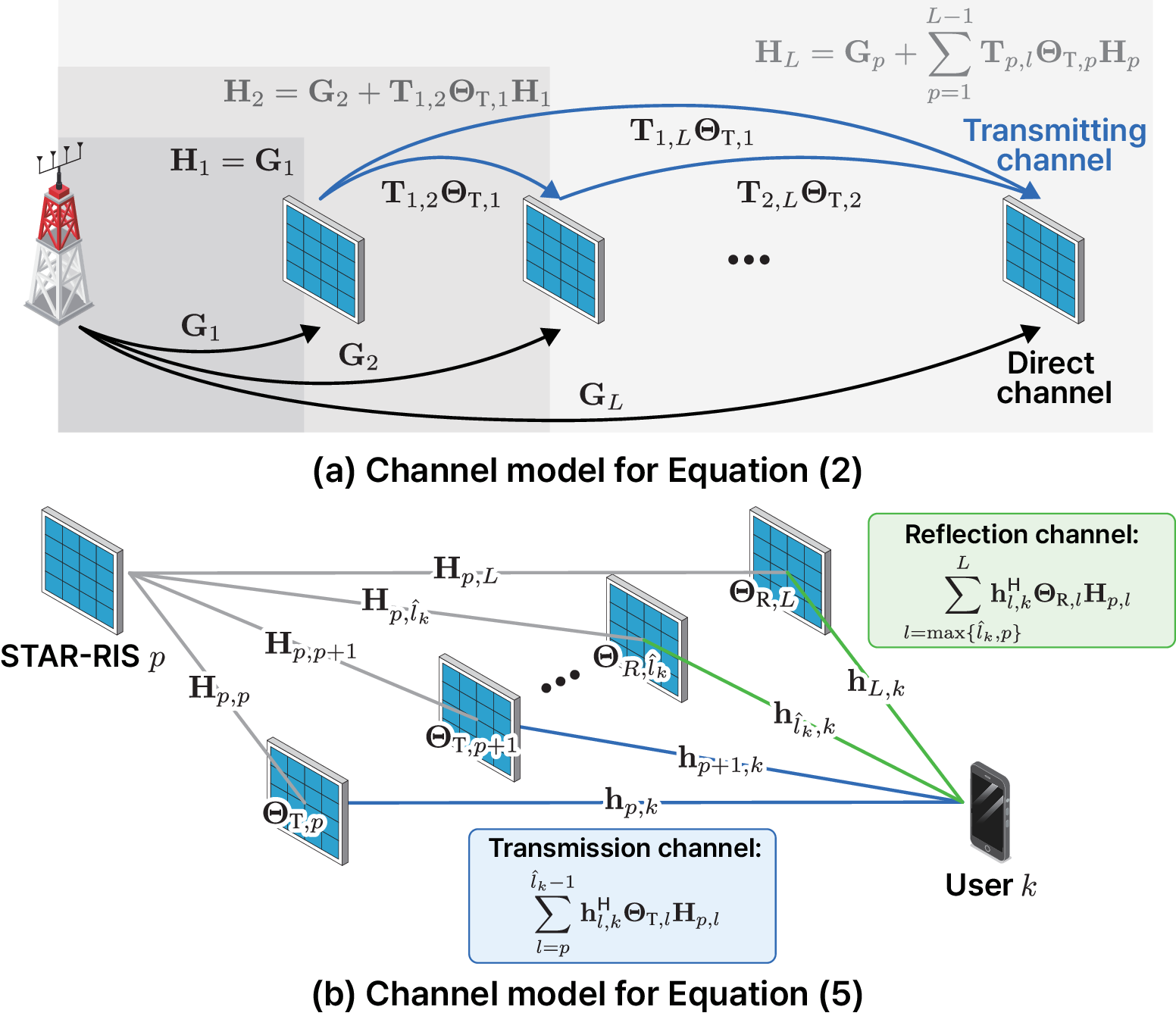}
    \vspace{-5pt}
    \caption{Illustration of the channel models. Channel models in \eqref{eq:H_p,l} and \eqref{eq:H_k} are similarly defined as in (a) and (b), respectively.}
    \label{fig:channel model}
    \vspace{-5pt}
\end{figure}

Fig.~\ref{fig:channel model} visually illustrates the main channel models of the proposed system. 
The sub-figures intuitively explain how the T\&R paths are accumulated to form the effective channels.
We define three channel models: i) from BS to STAR-RIS, $m \in \{ \text{T}, \text{R} \}$; ii) from STAR-RIS to STAR-RIS, $m \in \{ \text{T}, \text{R} \}$; and iii) from STAR-RIS to user, $\mathbf{h}_{l,k} \in \mathbb{C}^{M \times 1}$. 
We set index $S$ $(S \notin \mathcal{K})$ as the sensing target, i.e., $\mathbf{h}_{l,S}$ means the channel from STAR-RIS $l$ to the sensing target. 
We assume that all channels follow Rician fading. 

Let communication user $k$ be located in the transmission space of the $(\hat{l}_{k}-1)$-th STAR-RIS and the reflection space of the $\hat{l}_{k}$-th STAR-RIS.
The channel from the BS to STAR-RIS $l$ can be recursively defined as
\begin{align}
    \mathbf{H}_{l} = \mathbf{G}_p + \sum_{p=1}^{l-1} \mathbf{T}_{p,l} \boldsymbol{\Theta}_{\text{T},p} \mathbf{H}_p.
\end{align}
Additionally, the channel from STAR-RIS $p$ to STAR-RIS $l$ $(p \leq l)$ is given by
\begin{align}
    \mathbf{H}_{p,l} &=\begin{cases}
    \mathbf{I}_M, & \text{for} ~ p = l,\\
    \mathbf{T}_{p,l} \boldsymbol{\Theta}_{\text{T},p} + \sum\limits_{q=p+1}^{l-1} \mathbf{T}_{q,l} \boldsymbol{\Theta}_{\text{T},q} \mathbf{H}_{p,q}, & \text{for} ~ p<l.
  \end{cases}
  \label{eq:H_p,l}
\end{align}

The effective channel from the BS to user $k$ is denoted as 
\begin{align}
    \bar{\mathbf{h}}_{k}^\mathsf{H} &= \sum_{l=1}^{\hat{l}_k-1}  \mathbf{h}_{l,k}^\mathsf{H} \boldsymbol{\Theta}_{\text{T},l} \mathbf{H}_l + \sum_{l=\hat{l}_k}^{L}  \mathbf{h}_{l,k}^\mathsf{H} \boldsymbol{\Theta}_{\text{R},l} \mathbf{H}_l.
    \label{eq:H_k}
\end{align}
As in Fig.~\ref{fig:channel model}, the first term represents the effective channel through the transmission space, while the second term represents the effective channel through the reflection space.
Then, the effective channel from the STAR-RIS $p$ to user $k$ is 
\begin{align}
    \bar{\mathbf{h}}_{p,k}^\mathsf{H} &= \sum_{l=p}^{\hat{l}_k-1} \mathbf{h}_{l,k}^\mathsf{H} \boldsymbol{\Theta}_{\text{T},l} \mathbf{H}_{p,l} 
    +\hspace{-0.4cm} \sum_{l= \max \{ \hat{l}_k, p \} }^L \hspace{-0.5cm} \mathbf{h}_{l,k}^\mathsf{H} \boldsymbol{\Theta}_{\text{R},l} \mathbf{H}_{p,l}.
    \label{eq:H_p,k}
\end{align}
Then, the signal received at the $k$-th user is
\begin{align}
\label{eq:received_signal_comm_k}
    y_{k} &= \sum_{c \in \mathcal{K}} \bar{\mathbf{h}}_{k}^\mathsf{H} \mathbf{w}_{c} c_{c} + \bar{\mathbf{h}}_{k}^\mathsf{H} \mathbf{w}_S s + \sum_{p=1}^L \bar{\mathbf{h}}_{p,k}^\mathsf{H} \mathbf{v}_p + z_{k},
\end{align}
where $\mathbf{v}_p$ is the thermal noise induced by the inherent device noise of active elements in STAR-RIS and $z_{k}$ denotes the complex additive white Gaussian noise with distribution $\mathcal{CN}(0, \sigma_k^2)$. 
Then, the SINR of the $k$-th user reads
\begin{align}
    \gamma_{k} = \frac{\left| \bar{\mathbf{h}}_{k}^\mathsf{H} \mathbf{w}_{k} \right|^2}{\sum_{c\neq k} \left| \bar{\mathbf{h}}_{k}^\mathsf{H} \mathbf{w}_{c} \right|^2 + \sum_{p=1}^L \sigma_p^2 \| \bar{\mathbf{h}}_{p,k}^\mathsf{H}\|^2  + \sigma_{k}^2}.
\end{align}
Then, the achievable capacity of user $k$ is $r_{k} = \log (1+\gamma_{k})$.

The received signal at the BS can be represented by
\begin{align}
    &\mathbf{y}^{(\text{s})} = \overbrace{\rule{0pt}{0.377cm} \bar{\rho}_S {\bar{\mathbf{h}}_S} \bar{\mathbf{h}}_S^\mathsf{H}\mathbf{w}_S s}^{\text{Desired signal}} +
    \overbrace{\sum_{c \in \mathcal{K}} \bar{\rho}_S \bar{\mathbf{h}}_S \bar{\mathbf{h}}_{S}^\mathsf{H} \mathbf{w}_{c} c_{c}}^{\text{Communication signal}} \nonumber \\[-0.3cm]
    & + \underbrace{\bar{\rho}_S \bar{\mathbf{h}}_S \sum_{p=1}^L \bar{\mathbf{h}}_{p,S}^\mathsf{H} \mathbf{v}_p}_{\text{Thermal noise in forward path}} + \underbrace{\sum_{p=1}^L \mathbf{H}_{p}^\mathsf{H} \mathbf{v}_p}_{\text{Thermal noise in backward path}}  + \mathbf{z}_S,
\end{align}
where $\bar{\rho}_S$ represents the radar cross section (RCS) of the sensing target with $\mathbb{E}(|\bar{\rho}_S|^2) = \rho^2$. Note that the thermal noise reflected by the users can be ignored, since the inequality $\mathbf{H}_{p}^\mathsf{H} \mathbf{v}_p \gg \mathbf{H}_{p}^\mathsf{H} \mathbf{h}_{l,k} \mathbf{h}_{l,k}^\mathsf{H} \mathbf{v}_p $ is generally satisfied.
The sensing SINR at the BS is then given by \cite{Shuai23_TWC} 
\begin{align}
    \hspace{-0.22cm}
    \gamma_S = 
    \frac{\left| \bar{\mathbf{h}}_{S}^\mathsf{H} \mathbf{w}_S \right|^2}
    {\sum\limits_{c\in \mathcal{K}} \big| \bar{\mathbf{h}}_{S}^\mathsf{H} \mathbf{w}_{c} \big|^2 
    \! + \! \sum\limits_{p=1}^L \sigma_p^2 \big| \bar{\mathbf{h}}_{p,S}^\mathsf{H} \big|^2 
    \! + \! \sum\limits_{p=1}^L \frac{\sigma_p^2 \| \mathbf{H}_{p}^\mathsf{H}\|_\text{F}^2}{ \|\bar{\rho}_S\|^2 \left| \bar{\mathbf{h}}_{S} \right|^2}   
    \! + \! \sigma_S^2}, \!
\end{align}
where $\sigma_{S}^2 = \bar{\sigma}_S^2 / \big(|\bar{\rho}_S|^2 \left\| \bar{\mathbf{h}}_{S} \right\|^2\big)$. 

Since the sensing target may act as an eavesdropper, it attempts to extract communication data from the signals it receives. The SINR at the sensing target for intercepting the $k$-th user can be given by
\begin{align}
    \gamma_{\text{e},k} = \frac{\left| \bar{\mathbf{h}}_{S}^\mathsf{H} \mathbf{w}_{k} \right|^2}{\sum_{c\neq k} \left| \bar{\mathbf{h}}_{S}^\mathsf{H} \mathbf{w}_{c} \right|^2 + \sum_{p=1}^L \sigma_p^2 \| \bar{\mathbf{h}}_{p,S}^\mathsf{H}\|^2 + \sigma_\text{e}^2},
\end{align}
where $\sigma_\text{e}^2$ is the noise variance at the sensing target.

\subsection{Problem Formulation}
\begin{figure}[htbp]
    \centering
    \includegraphics[width=0.95\linewidth]{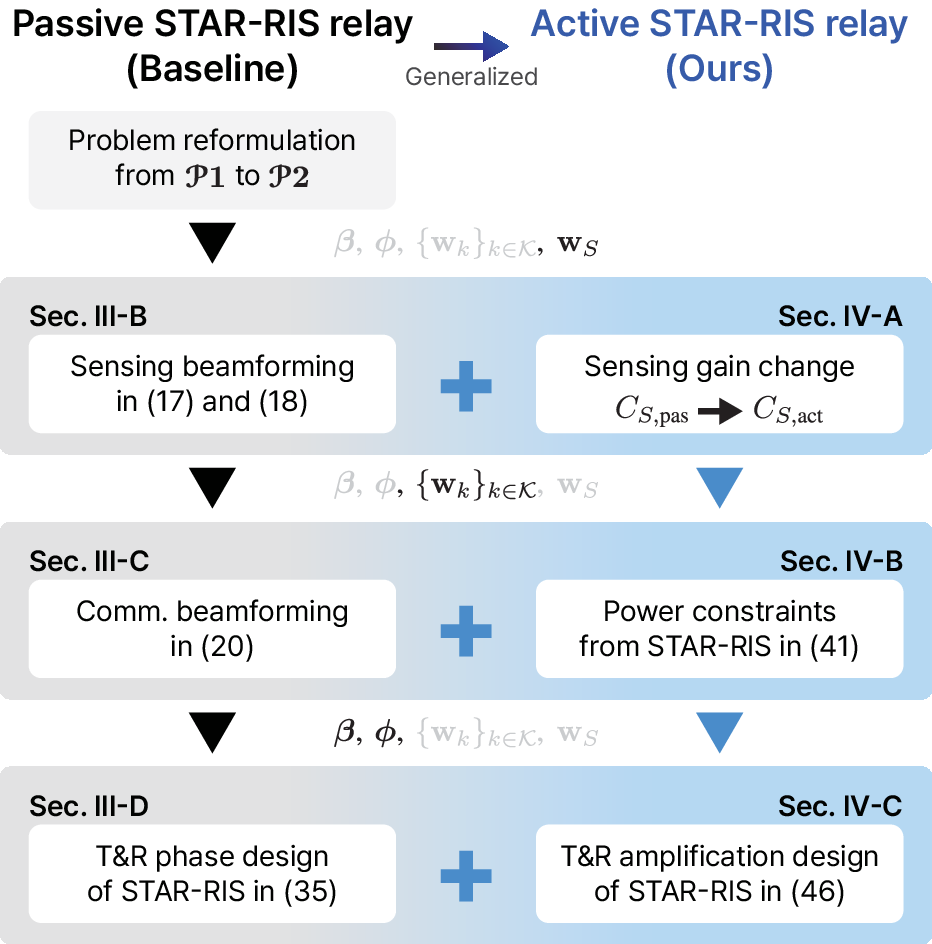}
    \caption{Visualization of the proposed beamforming design for passive and active STAR-RISs in Secs.~\ref{sec:Beamforming Design with Passive STAR-RIS} and~\ref{sec:Beamforming Design with Active STAR-RIS}. \textcolor[HTML]{C4C4C4}{Grey fonts} represent fixed variables.}
    \label{fig:solution_flowchart}
\end{figure}

We define the objective function as the weighted sum of data rate and sensing SINR under the constraints on sensing performance and power. By solving the expectation of the squared Euclidean norm of the radiated signals, the sum of T\&R power of the active RIS is
\begin{align}
    \hspace{-0.05cm} P_{A,l} &= \sum_{k \in \mathcal{K} \cup \{ S \}} \| \boldsymbol{\Theta}_{\text{T},l} \mathbf{H}_l \mathbf{w}_k \|^2 + \| \boldsymbol{\Theta}_{\text{T},l} \|_\text{F}^2 \sigma_l^2
    \nonumber \\
    & + \sum_{k \in \mathcal{K} \cup \{ S \}} \| \boldsymbol{\Theta}_{\text{R},l} \mathbf{H}_l \mathbf{w}_k \|^2 + \| \boldsymbol{\Theta}_{\text{R},l} \|_\text{F}^2 \sigma_l^2,
\end{align}
where $\sigma_l^2$ is the noise variance at STAR-RIS $l$.
Then, the optimization problem can be formulated as 
\begin{subequations}%
\label{eq:opt}
    \begin{alignat}{3}
        & \Problem{1}: && \max_{\boldsymbol{\beta},\,\boldsymbol{\phi},\,\{\mathbf{w}_k\}} && 
        \sum_{k\in\mathcal{K}} \alpha_k r_k,
        \label{subeq:opt_a}\\
        & && \mathrm{~~~~~s.t.~} && 
        \mathrm{tr}\Bigl(\hspace{-0.2cm}\sum_{k\in\mathcal{K}\cup\{S\}}\hspace{-0.2cm}\mathbf{w}_k\mathbf{w}_k^{\mathsf H}\Bigr)
        \le P_{\text{BS,max}},
        \label{subeq:opt_b}\\
        & && && P_{A,l} \le P_{\text{A},l,\text{max}},\; \forall l \in \mathcal{L}, 
        \label{subeq:opt_c}\\
        & && && 
        \gamma_{\text{e},k} \le \gamma_{\text{e},k,\text{req}}, \forall k\in\mathcal{K},
        \label{subeq:opt_d}\\
        & && && 
        \gamma_{S} \ge \gamma_{S,\text{req}},
        \label{subeq:opt_e}\\
        & && && 
        0 \le \phi_{\text{T},l,i},\,\phi_{\text{R},l,i} \le 2\pi,
         \forall l\in\mathcal{L},\, i\in\mathcal{M},
        \label{subeq:opt_f}
    \end{alignat}
\end{subequations}
where $\boldsymbol{\beta}$ and $\boldsymbol{\phi}$ are the sets including all amplitude $\beta_{m,l,i}$ and phase shift $\phi_{m,l,i}$ for $m \in \{ \text{T}, \text{R} \}$, $l\in \mathcal{L}$, and $i \in \mathcal{M}$, $\gamma_{S,\text{req}}$ and $\gamma_{\text{e},k,\text{req}}$ are the sensing SINR of the target and the maximum tolerable leakage of the communication information for user $k$ to the sensing target, respectively, $\mathcal{L} = \{ 1,\ldots, L \}$, and $\mathcal{M} = \{ 1,\ldots, M\}$.
Constraint \eqref{subeq:opt_b} and \eqref{subeq:opt_c} represent the power constraints on the BS transmit power and T\&R power of the active STAR-RIS $l$, respectively. Constraint \eqref{subeq:opt_d} indicates the security constraint for communication and Constraint \eqref{subeq:opt_e} is the performance constraint for sensing. Constraint \eqref{subeq:opt_f} indicates the corresponding bounds for the phase shift in $\boldsymbol{\theta}_{\text{T},l}$ and $\boldsymbol{\theta}_{\text{R},l}$.
We solve problem $\mathbf{\mathdutchcal{P}1}$ with passive and active STAR-RISs, as shown in Fig.~\ref{fig:solution_flowchart}. 

\section{Beamforming Design with Passive STAR-RISs}
\label{sec:Beamforming Design with Passive STAR-RIS}

We first optimize the beamforming of the passive STAR-RISs. Problem $\Problem{1}$ is then reformulated as 
\begin{subequations}%
\label{eq:opt_passive}
    \begin{alignat}{3}
        & \Problem{2}: && \max_{\boldsymbol{\beta},\,\boldsymbol{\phi},\,\{\mathbf{w}_k\}} && 
        \sum_{k\in\mathcal{K}} \alpha_k r_k,
        \label{subeq:opt_passive_a}\\[2pt]
        & && \mathrm{~~~~s.t.~} && 
        \beta_{\text{T},l,i}^{2} + \beta_{\text{R},l,i}^{2} = 1,
        \forall l\in\mathcal{L},\, i\in\mathcal{M},
        \label{subeq:opt_passive_b}\\[2pt]
        & && && 
        0 \le \beta_{\text{T},l,i},\,\beta_{\text{R},l,i} \le 1,
        \forall l\in\mathcal{L},\,i\in\mathcal{M},
        \label{subeq:opt_passive_c}\\
        & && && 
        \eqref{subeq:opt_b},~\eqref{subeq:opt_d},~\eqref{subeq:opt_e},
    \end{alignat}
\end{subequations}
where constraint \eqref{subeq:opt_passive_b} shows the law of energy conservation, and constraint \eqref{subeq:opt_passive_c} indicates the corresponding bounds for the amplitude and phase shift in $\boldsymbol{\theta}_{\text{T},l}$ and $\boldsymbol{\theta}_{\text{R},l}$. 
Problem $\Problem{2}$ is NP-hard due to the non-concave objective and coupled variables.
We introduce an AO algorithm that updates the transmit beamformers, and the T\&R phases and amplitudes.

Then, the following lemma is introduced to make the non-concave objective in $\Problem{2}$ tractable:
\lemma
Problem $\Problem{2}$ is equivalent to the below problem.
\begin{subequations}%
\label{eq:opt_tx_eqi}
    \begin{alignat}{3}
        & && \min_{\substack{\boldsymbol{\beta},\,\boldsymbol{\phi},\,\{\mathbf{w}_k\},\\\{u_k\},\,\{\xi_k\}}} && 
        \sum_{k\in\mathcal{K}} \alpha_k \bigl( \xi_k e_k - \log\xi_k \bigr),
        \label{subeq:opt_equi_a}\\[2pt]
        & && \mathrm{~~~~s.t.~} && 
        \eqref{subeq:opt_b},~\eqref{subeq:opt_d},~\eqref{subeq:opt_e},~\eqref{subeq:opt_passive_b},~\eqref{subeq:opt_passive_c},
        \label{opt_tx_eqi_b}\\[2pt]
        & && && 
        e_k =
        |u_k|^2 \biggl(
            \sum_{c\in\mathcal{K}\cup\{S\}} |\bar{\mathbf{h}}_k^{\mathsf H}\mathbf{w}_c|^2 + \sigma_k^2
        \biggr) \nonumber \\
        & && && ~~~~~~ 
        - 2\,\mathrm{Re}\bigl\{u_k\bar{\mathbf{h}}_k^{\mathsf H}\mathbf{w}_k\bigr\} + 1, 
         \forall k\in\mathcal{K},
        \label{subeq:opt_eqi_c}
    \end{alignat}
\end{subequations}

\proof 
The proof is given in Appendix \ref{proof:lemma_1}.

The objective function in \eqref{eq:opt_tx_eqi} is convex with respect to $\{ \mathbf{w}_k \}, \mathbf{w}_S, \{ u_k \}, \{ \xi_k \}$.  
We then decompose the overall optimization problem into three sub-problems: i) optimizing $\mathbf{w}_S$; ii) optimizing ${\mathbf{w}_k}, {u_k}, {\xi_k}$; and iii) optimizing the STAR coefficients ${\boldsymbol{\beta}, \boldsymbol{\phi}}$. These three groups are optimized iteratively until convergence.

\subsection{Solution for Sensing Transmit Beamformer}
With all variables ($\boldsymbol{\beta}$, $\boldsymbol{\phi}$, $\{ \mathbf{w}_k \}_{k \in \mathcal{K}}$, $\{ u_k \}_{k \in \mathcal{K}}$, $\{ \xi_k \}_{k \in \mathcal{K}}$) fixed except for $\mathbf{w}_S$, we can get the optimization problem with respect to the transmit beamformer for sensing as
\begin{subequations}%
\label{eq:opt_tx_eqi_sensing}
    \begin{alignat}{3}
        & && \min_{\mathbf{w}_S} && 
        \mathbf{w}_S^{\mathsf H} \mathbf{R}_{\text{int}} \mathbf{w}_S,
        \label{subeq:opt_tx_eqi_sensing_a}\\[2pt]
        & && \mathrm{~s.t.~} && 
        \eqref{subeq:opt_b},~
        |\bar{\mathbf{h}}_S^{\mathsf H} \mathbf{w}_S|^2 \ge C_{S,\text{pas}},
        \label{subeq:opt_tx_eqi_sensing_c}
    \end{alignat}
\end{subequations}
where the interference covariance matrix $\mathbf{R}_\text{int}$ is defined as $\mathbf{R}_\text{int} = \sum_{k \in \mathcal{K}} \alpha_k \xi_k |u_k|^2 \bar{\mathbf{h}}_k \bar{\mathbf{h}}_k^\mathsf{H}$, and $C_{S, \text{pas}}$ is given by
\begin{align}
    C_{S, \text{pas}} &= \max \{ \gamma_{S,\text{req}}(\sum\limits_{c\in \mathcal{K}} \left| \bar{\mathbf{h}}_{S}^\mathsf{H} \mathbf{w}_{c} \right|^2 + \sigma_S^2), \nonumber \\[-0.1cm]
    & \hspace{20pt} \frac{1}{\gamma_{\text{e},k,\text{req}}}\left| \bar{\mathbf{h}}_{S}^\mathsf{H} \mathbf{w}_{k} \right|^2 - \sum_{c\in \mathcal{K} \backslash \{k\}} \left| \bar{\mathbf{h}}_{S}^\mathsf{H} \mathbf{w}_{c} \right|^2 - \sigma_\text{e}^2 \}.
\end{align}
Note that constraints~\eqref{subeq:opt_d} and~\eqref{subeq:opt_e} both concern the minimum value of $\bar{\mathbf{h}}_S^\mathsf{H}\mathbf{w}_S$. Therefore, they can be reduced into the second term in \eqref{subeq:opt_tx_eqi_sensing_c}.
Solving this minimization problem is analogous to the classic MVDR beamforming problem with the power constraint. Consequently, the optimal solution can be shown to be
\begin{align}
\label{eq:bf_opt_sensing}
    \mathbf{w}_{S,\text{pas}} = \frac{\sqrt{C_{S, \text{pas}}} (\mathbf{R}_{\text{int}} + \epsilon \mathbf{I})^{-1} \bar{\mathbf{h}}_S}{\bar{\mathbf{h}}_S^\mathsf{H} (\mathbf{R}_{\text{int}} + \epsilon \mathbf{I})^{-1} \bar{\mathbf{h}}_S},
\end{align}
where $\epsilon$ is the Lagrange multiplier for the constraint \eqref{subeq:opt_b}.
Then, we can determine $\epsilon$ using the complementary slackness condition of constraints \eqref{subeq:opt_b}. 
Let $\mathbf{w}_{S,\text{pas}} (\epsilon)$ denote the right-hand side of \eqref{eq:bf_opt_sensing}. If $\mathbf{R}_{\text{int}}$ is invertible and $\mathrm{tr} (\mathbf{w}_{S, \text{pas}} (0) (\mathbf{w}_{S, \text{pas}} (0))^\mathsf{H}) \leq P_{\text{BS}, \text{max}} - \mathrm{tr} ( \sum_{k\in\mathcal{K}} \mathbf{w}_k \mathbf{w}_k^{\mathsf{H}})$, we have $\epsilon^{\text{opt}} = 0$. Otherwise, the constraint \eqref{subeq:opt_b} must hold with equality. 

\proposition For $\epsilon \ge 0$, the power of sensing beamformer $\Delta(\epsilon) = (\mathbf{w}_{S, \text{pas}} (\epsilon))^\mathsf{H} \mathbf{w}_{S, \text{pas}} (\epsilon)$ is a non-increasing function with respect to $\epsilon$, and has the lower bound of $\frac{C_{S, \text{pas}}}{\bar{\mathbf{h}}_S^\mathsf{H} \bar{\mathbf{h}}_S}$.
\proof Proof is given in Appendix \ref{proof:prop_1}.

\begin{table*}[!b]
\vspace{-0.3cm}
\hrule
\vspace{-0.2cm}
\begin{align}
\label{eq:lagrange}
\medskip
    L_{\text{pas}}( \{ \mathbf{w}_k\}, \lambda,&\{\mu_k\} )  = \sum_{k \in \mathcal{K}} \alpha_k \xi_k ( |u_k|^2 ( \sum_{c\in \mathcal{K} \cup \{ S \}} | \bar{\mathbf{h}}_k^\mathsf{H} \mathbf{w}_c |^2 + \sigma_k^2 ) - 2 \text{Re} \{ u_k \bar{\mathbf{h}}_k^\mathsf{H} \mathbf{w}_k \} + 1 ) + \lambda (\mathrm{tr} \Big( \sum_{k\in\mathcal{K} \cup \{ S \}} \hspace{-0.3cm}\mathbf{w}_k \mathbf{w}_k^\mathsf{H} \Big) - P_{\text{BS}, \text{max}}) \nonumber \\ 
    &~~~~~~~~~~ + \sum_{k \in \mathcal{K}} \mu_k (\left| \bar{\mathbf{h}}_{S}^\mathsf{H} \mathbf{w}_{k} \right|^2 - \gamma_{\text{e},k,\text{req}} (\sum_{c\neq k} \left| \bar{\mathbf{h}}_{S}^\mathsf{H} \mathbf{w}_{c} \right|^2 + \sigma_\text{e}^2)) - \mu_S ( |\bar{\mathbf{h}}_S^\mathsf{H} \mathbf{w}_S|^2 - \gamma_{S,\text{req}}(\sum\limits_{c\in \mathcal{K}} \left| \bar{\mathbf{h}}_{S}^\mathsf{H} \mathbf{w}_{c} \right|^2 + \sigma_S^2) ).
\protect
\end{align}
\end{table*}

From Proposition 1, we can conclude that when $\frac{C_{S, \text{pas}}}{\bar{\mathbf{h}}_S^\mathsf{H} \bar{\mathbf{h}}}_S \leq P_{\text{BS}, \text{max}} - \mathrm{tr} ( \sum_{k\in\mathcal{K}} \mathbf{w}_k \mathbf{w}_k^{\mathsf{H}})$, the optimal value of $\epsilon$ can be easily obtained via one-dimensional search algorithms. 
However, $\Delta(\epsilon)$ involves matrix inverses in both the numerator and denominator, resulting in a computational complexity of $\mathcal{O}(N^3)$ per evaluation.
We reduce the computational complexity by leveraging the eigenvalue decomposition of $\mathbf{R}_{\text{int}} = \mathbf{D} \boldsymbol{\Lambda} \mathbf{D}^\mathsf{H}$. Then, $\Delta(\epsilon)$ can be expressed as
\begin{align}
    \Delta(\epsilon) = C_{S, \text{pas}} \frac{ \sum_{i=1}^N |[\mathbf{g}_i|^2}{([\boldsymbol{\Lambda}]_{i,i} + \epsilon)^2} 
    \bigg( \sum_{i=1}^N \frac{|[\mathbf{g}]_i|^2}{([\boldsymbol{\Lambda}]_{i,i} + \epsilon)} \bigg)^{-2},
\end{align}
where $\mathbf{g} = \mathbf{D}^\mathsf{H} \bar{\mathbf{h}}_S$.
Then, the evaluation complexity of computing $\Delta(\epsilon)$ is significantly reduced from $\mathcal{O}(N^3)$ to $\mathcal{O}(N)$.

\subsection{Solution for Communication Transmit Beamformers}
\label{subsec:solution_tx_beamformer}
With the fixed $\boldsymbol{\beta}$, $\boldsymbol{\phi}$, and $\mathbf{w}_S$, problem \eqref{eq:opt_tx_eqi} becomes convex with respect to each of the optimization variables $\boldsymbol{\beta}$, $\boldsymbol{\phi}$, and $\{ \mathbf{w}_k \}$. We propose using the block coordinate descent method to solve the problem in \eqref{eq:opt_tx_eqi}. Specifically, we find the optimal solution for the problem by sequentially fixing two of the three variables and updating the third. The variable $u_k$ can be updated using the closed-form equation in \eqref{eq:receive_equalizer} presented in Appendix~\ref{proof:lemma_1}. Meanwhile, the update for $\{\xi_k\}$ is performed by setting $\xi_k = e_k^{-1}$, where $e_k$ is defined in \eqref{subeq:opt_eqi_c}.

To update $\{ \mathbf{w}_k \}$, we define the Lagrange function \eqref{eq:lagrange} at the bottom of this page.
The first-order optimality condition of the Lagrange function with respect to $\mathbf{w}_{k}$ for $k \in \mathcal{K}$ yields
\begin{align}
\label{w_k_lagrange}
    \mathbf{w}_{k, \text{pas}}= \Big(\sum_{\ell \in \mathcal{K}} \alpha_\ell \xi_\ell & |u_\ell|^2 \bar{\mathbf{h}}_\ell \bar{\mathbf{h}}_\ell^\mathsf{H} + \lambda \mathbf{I} \\[-0.3cm]
    &~~~ -\hspace{-0.2cm} \sum_{\ell \in \mathcal{K} \cup \{ S \} } \hspace{-0.2cm} \mu_\ell \bar{\gamma}_{\ell,\text{req}}^{(k)} \bar{\mathbf{h}}_S \bar{\mathbf{h}}_S^\mathsf{H}\Big)^{-1} \alpha_k \xi_k u_k^{*} \bar{\mathbf{h}}_k,  \nonumber
\end{align}
where $\bar{\gamma}_{\text{e},\ell,\text{req}}^{(k)}$ and $\bar{\gamma}_{S,\text{req}}^{(k)}$ are defined as
\begin{align}
\label{eq:gamma_cases}
    \bar{\gamma}_{\ell,\text{req}}^{(k)} &=\begin{cases}
    -1, & \text{for} ~ \ell = k,\\
    \gamma_{\text{e},\ell,\text{req}}, & \text{for} ~ \ell \neq k,S,\\
    -\gamma_{S,\text{req}}, & \text{for} ~ \ell = S.
  \end{cases}
\end{align}
By the complementary slackness condition of \eqref{subeq:opt_b}, \eqref{subeq:opt_d}, and \eqref{subeq:opt_e}, we have two cases according to the value of the Lagrange multipliers, i.e., $\lambda$ and $\mu_k$.

\subsubsection{Solution for $\lambda$} With the fixed $\boldsymbol{\mu} = \{ \mu_1, \ldots, \mu_K, \mu_S \}$, we let $\mathbf{w}_{k,\text{pas}} (\lambda, \boldsymbol{\mu})$ denote the right-hand side of \eqref{w_k_lagrange}. If $\sum_{\ell \in \mathcal{K} \cup \{ S \}} \alpha_\ell \xi_\ell |u_\ell|^2 \bar{\mathbf{h}}_\ell \bar{\mathbf{h}}_\ell^\mathsf{H} - \sum_{\ell \in \mathcal{K}} \mu_\ell \bar{\gamma}_{\text{e},\ell,\text{req}}^{(k)} \bar{\mathbf{h}}_S \bar{\mathbf{h}}_S^\mathsf{H}$ is invertible and $\mathrm{tr} ( \sum_{k\in\mathcal{K} \cup \{ S \}} \mathbf{w}_{k,\text{pas}}(0, \boldsymbol{\mu}) (\mathbf{w}_{k, \text{pas}} (0, \boldsymbol{\mu}))^\mathsf{H} ) \leq P_{\text{BS}, \text{max}}$, we have $\lambda^{\text{opt}} = 0$. Otherwise, the constraint \eqref{subeq:opt_b} must hold with equality. Let $\mathbf{D}_k \boldsymbol{\Lambda}_k \mathbf{D}_k^\mathsf{H}$ represent the eigenvalue decomposition of $\sum_{\ell \in \mathcal{K} \cup \{ S \}} \alpha_\ell \xi_\ell |u_\ell|^2 \bar{\mathbf{h}}_\ell \bar{\mathbf{h}}_\ell^\mathsf{H} - \sum_{\ell \in \mathcal{K}} \mu_\ell \bar{\gamma}_{\text{e},\ell,\text{req}}^{(k)} \bar{\mathbf{h}}_S \bar{\mathbf{h}}_S^\mathsf{H}$. Then, the following equation holds:
\begin{align}
\label{eq:eigenvalue_decom}
    \sum_{k \in \mathcal{K} } \mathrm{tr} \big( (\boldsymbol{\Lambda}_k + \lambda \mathbf{I} )^{-2} \boldsymbol{\Upsilon}_k\big) = P_{\text{BS}, \text{max}} - \mathbf{w}_S^\mathsf{H} \mathbf{w}_S,
\end{align}
where $\boldsymbol{\Upsilon}_k = \mathbf{D}_k^\mathsf{H} ( |\alpha_k|^2 |\xi_k|^2 |u_k|^2 \bar{\mathbf{h}}_k \bar{\mathbf{h}}_k^\mathsf{H} ) \mathbf{D}_k$ and $P_S = \mathbf{w}_S^\mathsf{H} \mathbf{w}_S$. Then, the equation \eqref{eq:eigenvalue_decom} is equivalent to
\begin{align}
    \sum_{k \in \mathcal{K} } \sum_{i=1}^{N} \frac{[\boldsymbol{\Upsilon}_k]_{ii}}{( [\boldsymbol{\Lambda}_k]_{ii} + \lambda )^2} = P_{\text{BS}, \text{max}} - \mathbf{w}_S^\mathsf{H} \mathbf{w}_S.
\end{align}
With the fixed $\boldsymbol{\mu}$, the optimal value of $\lambda$ can be readily obtained via one-dimensional search algorithms.

\subsubsection{Solution for $\boldsymbol{\mu}$} 
The Lagrange multiplier $\mu_k$ appears as an inverse function of $\mathbf{w}_{k,\text{pas}}$, making the closed-form solution intractable.
We instead fix all other multipliers and optimize $\mu_k$ iteratively, repeating this until convergence.
The iteration is divided into two sub-processes corresponding to the constraints \eqref{subeq:opt_d} and \eqref{subeq:opt_e}.

\paragraph{Constraint \eqref{subeq:opt_d}}
With the fixed $\lambda$ and $\bm{\mu}_{-k}=\bm{\mu}\backslash\{k\}$, we define $\mathbf{E}_{k,m} = \sum_{\ell \in \mathcal{K} \cup \{ S \}} \alpha_\ell \xi_\ell |u_\ell|^2 \bar{\mathbf{h}}_\ell \bar{\mathbf{h}}_\ell^\mathsf{H} + \lambda \mathbf{I} $ $- \sum_{\ell \neq k} \mu_\ell \bar{\gamma}_{\ell,\text{req}}^{(m)} \bar{\mathbf{h}}_S \bar{\mathbf{h}}_S^\mathsf{H}$. 
If $\mathbf{E}_{k,m} - \mu_k \bar{\gamma}_{k,\text{req}}^{(m)} \bar{\mathbf{h}}_S \bar{\mathbf{h}}_S^\mathsf{H}$ is invertible and $\gamma_{\text{e},k} \leq \gamma_{\text{e},k,\text{req}}$ with $\mathbf{w}_{k, \text{pas}} (\lambda, \bm{\mu})$ for $\mu_k=0$, we have $\mu_k^{\text{opt}} =0$. Otherwise, the constraint \eqref{subeq:opt_c} should be satisfied with equality, which leads to $|\bar{\mathbf{h}}_S^\mathsf{H} \mathbf{w}_{m, \text{pas}}| = |\bar{\alpha}_m| |\bar{\mathbf{h}}_S^\mathsf{H} (\mathbf{E}_{k,m} - \mu_k \bar{\gamma}_{k,\text{req}}^{(m)} \bar{\mathbf{h}}_S \bar{\mathbf{h}}_S^\mathsf{H} )^{-1} \bar{\mathbf{h}}_m|$ where $\bar{\alpha}_m = \alpha_m \xi_m u_m$. By the Sherman-Morrison formula, the following equation holds:
\begin{align}
\label{eq:sherman-morrison}
    (\mathbf{E}_{k,m} - \mu_k \bar{\gamma}_{k,\text{req}}^{(m)} &\bar{\mathbf{h}}_S \bar{\mathbf{h}}_S^\mathsf{H} )^{-1} \nonumber\\ 
    &= \mathbf{E}_{k,m}^{-1} + \frac{\mu_k \bar{\gamma}_{k,\text{req}}^{(m)} \mathbf{E}_{k,m}^{-1} \bar{\mathbf{h}}_S \bar{\mathbf{h}}_S^\mathsf{H} \mathbf{E}_{k,m}^{-1} }{1-\mu_k \bar{\gamma}_{k,\text{req}}^{(m)} \bar{\mathbf{h}}_S^\mathsf{H} \mathbf{E}_{k,m}^{-1} \bar{\mathbf{h}}_S }.
\end{align}
Inserting \eqref{eq:sherman-morrison} into $\bar{\mathbf{h}}_S^\mathsf{H} \mathbf{w}_{m, \text{pas}}$ and simplifying yields
\begin{align}
\label{eq:hswm_abs}
    |\bar{\mathbf{h}}_S^\mathsf{H} \mathbf{w}_{m, \text{pas}}|^2 = |\bar{\alpha}_m|^2 |g_{k,m}|^2 |\frac{1}{1-\mu_k \bar{\gamma}_{k,\text{req}}^{(m)} \kappa_{k,m} }|^2,
\end{align}
where $g_{k,m} = \bar{\mathbf{h}}_S^\mathsf{H} \mathbf{E}_{k,m}^{-1} \bar{\mathbf{h}}_m$ and $\kappa_{k,m} = \bar{\mathbf{h}}_S^\mathsf{H} \mathbf{E}_{k,m}^{-1} \bar{\mathbf{h}}_S$.

Plugging \eqref{eq:hswm_abs} into \eqref{subeq:opt_d} simplifies it as
\begin{align}
\label{eq:final_mu}
    \hspace{-0.1cm} \sigma_\text{e}^2 + |\bar{\mathbf{h}}_S^\mathsf{H} \mathbf{w}_{S, \text{pas}}|^2 =&
    \frac{1}{\gamma_{\text{e},k,\text{req}}} |\bar{\alpha}_k|^2 |g_{k,k}|^2 |\frac{1}{1+\mu_k \kappa_{k,k} }|^2 \\
    & -\hspace{-0.2cm} \sum_{\ell \in \mathcal{K} \backslash \{ k \} } \hspace{-0.2cm} |\bar{\alpha}_\ell|^2 |g_{k,\ell}|^2 |\frac{1}{1-\mu_k \gamma_{\text{e}, k,\text{req}} \kappa_{k,\ell} }|^2. \nonumber 
\end{align}
We introduce the following proposition to check the existence of a nonnegative real solution to \eqref{eq:final_mu}, defining the right-hand side of \eqref{eq:final_mu} as a function $f(\mu_k)$.

\proposition 
\label{prop:mu_existence}
In the range of $(0, 1/(\gamma_{\text{e},k,\text{req}} \kappa_{k,l})]$, there exists a unique non-negative real solution for equation \eqref{eq:final_mu}.

\proof The proof is given in Appendix \ref{proof:prop_2}.

From Proposition \ref{prop:mu_existence}, the unique solution can be obtained via one-dimensional search algorithms.

\paragraph{Constraint \eqref{subeq:opt_e}}
Similar to the constraint \eqref{subeq:opt_d} case, if $\gamma_{S} \geq \gamma_{S,\text{req}}$ with $\mathbf{w}_{k, \text{pas}} (\lambda, \{ \mu_1, \ldots, \mu_K, 0 \})$, we have $\mu_S^{\text{opt}} =0$. Otherwise, the constraint \eqref{subeq:opt_c} and/or \eqref{subeq:opt_d} should be satisfied with equality. Through the same mathematical derivation as before, the following expression can be obtained:
\begin{align}
\label{eq:final_mu_S}
    \hspace{-0.18cm}\sum_{\ell \in \mathcal{K} } |\bar{\alpha}_\ell|^2 |g_{S,\ell}|^2 |\frac{1}{1-\mu_S \kappa_{S,\ell} }|^2 = \frac{1}{\gamma_{S,\text{req}}}|\bar{\mathbf{h}}_S^\mathsf{H} \mathbf{w}_{S, \text{pas}}|^2 - \sigma_{S}^2.
    \hspace{-0.1cm}
\end{align}
In the same manner as Proposition \ref{prop:mu_existence}, it can be proven that a unique non-negative real solution exists within the interval $(0, 1/\kappa_{S,S})$. Furthermore, the solution can be obtained using a one-dimensional search algorithm.

\subsection{Solution for Amplitude and Phase Shift Coefficients}
\label{subsec:passive_channel_opt}
Now, we fix the group of variables $\{ \{\mathbf{w}_k \}, \{u_k\}, \{ \xi_k \} \}$ and optimize the amplitude and phase shift $\{ \{ \mathbf{B} \}, \{ \boldsymbol{\Phi} \} \}$ of the STAR-RIS. Unfortunately, due to the coupled formulation of $\boldsymbol{\Theta}$ in $\bar{\mathbf{h}}_k^\mathsf{H}$, it is intractable to directly solve the optimization problem in \eqref{eq:opt_tx_eqi}.
We implement an iterative algorithm to efficiently solve the optimization problem with low complexity by individually updating the T\&R matrices of each STAR-RIS.
For each STAR-RIS $l$, we fix the T\&R matrices of all other STAR-RISs and optimize only those of STAR-RIS $l$.

With the fixed T\&R coefficient matrices for STAR-RIS $1,\ldots, q-1, q+1, \ldots, L$, we first represent the channel from the BS to STAR-RIS $l$ $(l\geq q)$ with respect to $\boldsymbol{\Theta}_{\text{T}, q}$ as
\begin{align}
    \mathbf{H}_{l} = \tilde{\mathbf{T}}_{q,l} \boldsymbol{\Theta}_{\text{T}, q} \mathbf{H}_{q} + (\mathbf{H}_l - \mathbf{H}_{q,l} \mathbf{H}_{q}),
\end{align}
where $\tilde{\mathbf{T}}_{q,l}$ is given by
\begin{align}
    \tilde{\mathbf{T}}_{q,l} = \begin{cases}
    \boldsymbol{\Theta}_{\text{T},q}^{-1}, ~\text{if} ~ q = l \\
    \mathbf{T}_{q,l} + \sum_{p=q+1}^{l-1} \mathbf{T}_{p,l} \boldsymbol{\Theta}_{\text{T},p} \tilde{\mathbf{T}}_{q,p}, ~\text{if} ~ q < l. \end{cases}
\end{align}
Subsequently, the effective channel from the BS to user $k$ with respect to $\boldsymbol{\theta}_{\text{T}, q}$ and $\boldsymbol{\theta}_{\text{R}, q}$ can be reformulated by
\begin{align}
    \hspace{-0.2cm}
    \mathbf{\bar{h}}_k^\mathsf{H} =  \bar{\boldsymbol{\theta}}_q^\mathsf{T} \bar{\mathbf{F}}_{k,q} + \mathbf{b}_{k,q}^\mathsf{H} = \boldsymbol{\theta}_{\text{T},q}^\mathsf{T} \mathbf{F}_{\text{T},k, q} + \boldsymbol{\theta}_{\text{R},q}^\mathsf{T} \mathbf{F}_{\text{R},k, q} + \mathbf{b}_{k,q}^\mathsf{H},
\end{align}
where $\bar{\boldsymbol{\theta}}_q^\mathsf{T} = [ \boldsymbol{\theta}_{\text{T},q}^\mathsf{T},  \boldsymbol{\theta}_{\text{R},q}^\mathsf{T}]$, $\bar{\mathbf{F}}_{k,q} = [\mathbf{F}_{\text{T},k,q}^\mathsf{T}, \mathbf{F}_{\text{R},k,q}^\mathsf{T}]^\mathsf{T}$, $\boldsymbol{\theta}_{m,q}^\mathsf{T} = \boldsymbol{\beta}_{m,q}^\mathsf{T} \circ \boldsymbol{\phi}_{m,q}^\mathsf{T}$, $\mathbf{F}_{\text{T},k,q}$, $\mathbf{F}_{\text{R},k,q}$, and $\mathbf{b}_{k,q}$ are defined as
\begin{align}
    \vspace{-0.1cm}
    \mathbf{F}_{\text{T},k,q} &= \sum_{l=q}^{\hat{l}_k-1} \mathrm{diag} (\mathbf{h}_{l,k}^{\mathsf{H}} \boldsymbol{\Theta}_{\text{T},l} \tilde{\mathbf{T}}_{q,l} ) \mathbf{H}_q \nonumber \\[-0.1cm]
    & \hspace{0.15cm}+ 
    \hspace{-0.75cm}
    \sum_{l=\max \{ \hat{l}_k,q+1 \} }^L \hspace{-0.7cm} \mathrm{diag} (\mathbf{h}_{l,k}^{\mathsf{H}} \boldsymbol{\Theta}_{\text{R},l} \tilde{\mathbf{T}}_{q,l} ) \mathbf{H}_q, \\
    \mathbf{F}_{\text{R},k,q} &= \begin{cases}
    \mathrm{diag}( \mathbf{h}_{q,k}^{\mathsf{H}} ) \mathbf{H}_q, ~\text{if} ~ q \geq \hat{l}_k, \\
    \mathbf{0}, \hspace{55pt} \text{otherwise}, \end{cases}
\end{align}
\vspace{-0.3cm}
\begin{align}
    \!\! & \mathbf{b}_{k,q}^\mathsf{H} = \hspace{-0.4cm}\sum_{l=1}^{\min \{ q-1, \hat{l}_k-1 \}} \hspace{-0.7cm}\mathbf{h}_{l,k}^\mathsf{H} \boldsymbol{\Theta}_{\text{T},l} \mathbf{H}_l + \sum_{l=\hat{l}_k}^{q-1} \mathbf{h}_{l,k}^\mathsf{H} \boldsymbol{\Theta}_{\text{R},l} \mathbf{H}_l \\[-0.3cm]
    & \hspace{-0.1cm} + \!\! \sum_{l=q+1}^{\hat{l}_k-1} \mathbf{h}_{l,k}^{\mathsf{H}} \boldsymbol{\Theta}_{\text{T},l} (\mathbf{H}_l \hspace{-2pt} - \hspace{-2pt} \mathbf{H}_{q,l} \mathbf{H}_q)
    + \hspace{-0.7cm} \sum_{l=\max \{ \hat{l}_k,q+1 \} }^L \hspace{-0.7cm} \mathbf{h}_{l,k}^{\mathsf{H}} \boldsymbol{\Theta}_{\text{R},l} (\mathbf{H}_l \hspace{-2pt} - \hspace{-2pt} \mathbf{H}_{q,l} \mathbf{H}_q). \nonumber
\end{align}
After omitting the constant terms in the objective function, we revisit the optimization problem in \eqref{eq:opt_tx_eqi} with respect to $\bar{\boldsymbol{\theta}}_q$ by
\begin{subequations}%
\label{eq:reform_opt_ris}
    \begin{alignat}{3}
        \hspace{-0.05cm}
        & && \min_{\bar{\boldsymbol{\theta}}_{q}} && 
        \sum_{k\in\mathcal{K}} \alpha_k \xi_k e_k, 
        \label{subeq:opt_ris_a} \\ 
        & && \mathrm{~s.t.~} && 
        [\mathbf{I},\,\mathbf{I}]\bigl( \bar{\boldsymbol{\theta}}_q \circ \bar{\boldsymbol{\theta}}_q^{*} \bigr) 
        = \mathbf{1}, 
        \label{subeq:opt_ris_b} \\ 
        & && && 
        \frac{1}{\gamma_{\text{e},k,\text{req}}} 
        \bigl( \bar{\boldsymbol{\theta}}_q^{\mathsf T}\bar{\mathbf{F}}_{S,q} + \mathbf{b}_{S,q}^{\mathsf H} \bigr) 
        \mathbf{W}_k 
        \bigl( \bar{\mathbf{F}}_{S,q}^{\mathsf H}\bar{\boldsymbol{\theta}}_q^{*} + \mathbf{b}_{S,q} \bigr)  \label{subeq:opt_ris_c} \\[-0.1cm]
        & && && ~~~
        \le 
        \bigl( \bar{\boldsymbol{\theta}}_q^{\mathsf T}\bar{\mathbf{F}}_{S,q} + \mathbf{b}_{S,q}^{\mathsf H} \bigr) 
        \hspace{-0.5cm} \sum_{c\in\mathcal{K}\cup\{S\}\setminus\{k\}} \hspace{-0.6cm} \mathbf{W}_c 
        \bigl( \bar{\mathbf{F}}_{S,q}^{\mathsf H}\bar{\boldsymbol{\theta}}_q^{*} + \mathbf{b}_{S,q} \bigr) 
        + \sigma_{\text{e}}^2, 
         \nonumber \\[0.05cm]
        & && && 
        \frac{1}{\gamma_{S,\text{req}}} 
        \bigl( \bar{\boldsymbol{\theta}}_q^{\mathsf T}\bar{\mathbf{F}}_{S,q} + \mathbf{b}_{S,q}^{\mathsf H} \bigr) 
        \mathbf{W}_S 
        \bigl( \bar{\mathbf{F}}_{S,q}^{\mathsf H}\bar{\boldsymbol{\theta}}_q^{*} + \mathbf{b}_{S,q} \bigr)   \label{subeq:opt_ris_d} \\[-0.1cm]
        & && && ~~~
        \ge 
        \bigl( \bar{\boldsymbol{\theta}}_q^{\mathsf T}\bar{\mathbf{F}}_{S,q} + \mathbf{b}_{S,q}^{\mathsf H} \bigr) 
        \sum_{c\in\mathcal{K}} \mathbf{W}_c 
        \bigl( \bar{\mathbf{F}}_{S,q}^{\mathsf H}\bar{\boldsymbol{\theta}}_q^{*} + \mathbf{b}_{S,q} \bigr) 
        + \sigma_S^2, 
         \nonumber \\[0.05cm]
        & && && 
        e_k =
        |u_k|^2\bigl( 
            \bar{\boldsymbol{\theta}}_q^{\mathsf T}\bar{\mathbf{F}}_{k,q}\bar{\mathbf{W}}\bar{\mathbf{F}}_{k,q}^{\mathsf H}\bar{\boldsymbol{\theta}}_q^{*} 
            + \bar{\boldsymbol{\theta}}_q^{\mathsf T}\bar{\mathbf{F}}_{k,q}\bar{\mathbf{W}}\mathbf{b}_{k,q} 
        \label{subeq:opt_ris_e} \\
        & && &&
            + \sigma_k^2
            + \mathbf{b}_{k,q}^{\mathsf H}\bar{\mathbf{W}}\bar{\mathbf{F}}_{k,q}^{\mathsf H}\bar{\boldsymbol{\theta}}_q^{*} 
        \bigr)
        \hspace{-1pt} - \hspace{-1pt} u_k \bar{\boldsymbol{\theta}}_q^{\mathsf T}\bar{\mathbf{F}}_{k,q}\mathbf{w}_k 
        - u_k^{*}\mathbf{w}_k^{\mathsf H}\bar{\mathbf{F}}_{k,q}^{\mathsf H}\bar{\boldsymbol{\theta}}_q^{*},
        \nonumber
    \end{alignat}
\end{subequations}
where $\mathbf{W}_c = \mathbf{w}_c \mathbf{w}_c^\mathsf{H}$ and $\bar{\mathbf{W}} = \sum_{c\in \mathcal{K} \cup \{ S \}} \mathbf{w}_c\mathbf{w}_c^\mathsf{H}$. The optimization problem \eqref{eq:reform_opt_ris} is non-convex due to the constant modulus constraint in \eqref{subeq:opt_ris_b}. Thus, we transform the optimization problem \eqref{eq:reform_opt_ris} into a SDP by introducing $\tilde{\boldsymbol{\Theta}}_q = \tilde{\boldsymbol{\theta}}_q^{*} \tilde{\boldsymbol{\theta}}_q^\mathsf{T}$ and $\tilde{\boldsymbol{\theta}}_q = [\bar{\boldsymbol{\theta}}_q^T, 1]^T$, which is given as
\begin{subequations}%
\label{eq:opt_ris}
    \begin{alignat}{3}
        & && \min_{\tilde{\boldsymbol{\Theta}}_q} && 
        \mathrm{tr}\bigl( \tilde{\boldsymbol{\Theta}}_q \mathbf{J}_q \bigr), 
        \\ 
        & && \mathrm{~~s.t.~} && 
        [\tilde{\boldsymbol{\Theta}}_q]_{i,i} + [\tilde{\boldsymbol{\Theta}}_q]_{M+i,M+i} = 1, \; i\in\mathcal{M}, 
        \label{subeq:reform_opt_ris_b} \\ 
        & && && 
        \mathrm{tr}\bigl( \tilde{\boldsymbol{\Theta}}_q \mathbf{M}_{k,q} \bigr) 
        \le \sigma_{\text{e}}^2, \; k\in\mathcal{K}, 
        \label{subeq:reform_opt_ris_c} \\ 
        & && && 
        \mathrm{tr}\bigl( \tilde{\boldsymbol{\Theta}}_q \mathbf{M}_{S,q} \bigr) 
        \ge \frac{\bar{\sigma}_S^2}{
            |\rho_S|^2 \mathrm{tr}\bigl( \tilde{\boldsymbol{\Theta}}_q \mathbf{N}_q \bigr) }, 
        \label{subeq:reform_opt_ris_d} \\
        & && && 
        [\tilde{\boldsymbol{\Theta}}_q]_{2M+1,2M+1} = 1, 
        \label{subeq:reform_opt_ris_e} \\ 
        & && && 
        \tilde{\boldsymbol{\Theta}}_q = \tilde{\boldsymbol{\Theta}}_q^{\mathsf H}, \; 
        \tilde{\boldsymbol{\Theta}}_q \succeq 0, 
        \label{subeq:reform_opt_ris_f} \\ 
        & && && 
        \mathrm{rank}\bigl( \tilde{\boldsymbol{\Theta}}_q \bigr) = 1,
        \label{subeq:reform_opt_ris_g}
    \end{alignat}
\end{subequations}
where $\mathbf{J}_q$, $\mathbf{M}_k$, and $\mathbf{N}_k$ are defined as
\begin{align}
    \mathbf{J}_{q} &= \sum_{k \in \mathcal{K}} \alpha_k \xi_k \\[-0.05cm]
    & \hspace{-6pt} \times
    \begin{pmatrix}
        |u_k|^2 \bar{\mathbf{F}}_{k,q} \bar{\mathbf{W}} \bar{\mathbf{F}}_{k,q}^\mathsf{H}
        & u_k \bar{\mathbf{F}}_{k,q} (u_k^{*}  \bar{\mathbf{W}} \mathbf{b}_{k,q}  - \mathbf{w}_k) \\
        (u_k \mathbf{b}_{k,q}^\mathsf{H} \bar{\mathbf{W}} - \mathbf{w}_k^\mathsf{H}) u_k^* \bar{\mathbf{F}}_{k,q}^\mathsf{H} & |u_k|^2 \sigma_k^2
    \end{pmatrix}, \nonumber \\[0.15cm]
    & \mathbf{M}_{k,q} =
    \begin{pmatrix}
        \bar{\mathbf{F}}_{S,q} \bar{\mathbf{W}}_{-k} \bar{\mathbf{F}}_{S,q}^\mathsf{H} & \bar{\mathbf{F}}_{S,q} \bar{\mathbf{W}}_{-k} \mathbf{b}_{S,q} \\[0.15cm]
        \mathbf{b}_{S,q}^\mathsf{H} \bar{\mathbf{W}}_{-k} \bar{\mathbf{F}}_{S,q}^\mathsf{H} & \mathbf{b}_{S,q}^\mathsf{H} \bar{\mathbf{W}}_{-k} \mathbf{b}_{S,q}
        \end{pmatrix}, \\[0.1cm]
    & \mathbf{N}_q =
    \begin{pmatrix}
        \bar{\mathbf{F}}_{S,q} \bar{\mathbf{F}}_{S,q}^\mathsf{H} & \bar{\mathbf{F}}_{S,q} \mathbf{b}_{S,q} \\
        \mathbf{b}_{S,q}^\mathsf{H} \bar{\mathbf{F}}_{S,q}^\mathsf{H} & \mathbf{b}_{S,q}^\mathsf{H} \mathbf{b}_{S,q}
        \end{pmatrix},
\end{align}
and $\bar{\mathbf{W}}_{-k} = \frac{1 + \gamma_{\text{e},k,\text{req}}}{\gamma_{\text{e},k,\text{req}}} \mathbf{W}_k - \bar{\mathbf{W}}$ for $k \in \mathcal{K}$ and $\bar{\mathbf{W}}_{-S} = \frac{1 + \gamma_{S,\text{req}}}{\gamma_{S,\text{req}}} \mathbf{W}_S - \bar{\mathbf{W}}$.
By omitting rank one constraint in \eqref{subeq:reform_opt_ris_g}, the optimization problem \eqref{eq:opt_ris} becomes convex with respect to $\boldsymbol{\Theta}_q$, which can be solved by standard SDP
optimization techniques with the CVX tool.

\section{Beamforming Design with Active STAR-RISs}
\label{sec:Beamforming Design with Active STAR-RIS}

In this section, we turn our attention to beamforming optimization problem with active STAR-RIS in $\Problem{1}$ where amplitudes of the STAR-RIS are not limited by unit modulus constraint. By Lemma 1, we can formulate the following optimization problem with active STAR-RIS:
\begin{subequations}%
\label{eq:opt_tx_eqi_act}
    \begin{alignat}{3}
        & \Problem{3}: && \min_{\substack{\boldsymbol{\beta},\,\boldsymbol{\phi},\,\{\mathbf{w}_k\},\\\{u_k\},\,\{\xi_k\}}} && 
        \sum_{k\in\mathcal{K}} \alpha_k \bigl( \xi_k e_k - \log\xi_k \bigr), 
        \\ 
        & && \mathrm{~~~~~s.t.} && 
        \eqref{subeq:opt_b} - \eqref{subeq:opt_e}, 
        \\ 
        & && && 
        e_k =
        |u_k|^2 
        \big(
            \sum_{c\in\mathcal{K}\cup\{S\}} \hspace{-0.2cm} \bigl| \bar{\mathbf{h}}_k^{\mathsf H} \mathbf{w}_c \bigr|^2 
            + \tilde{\sigma}_k^2 
        \big) \nonumber \\ 
        & && && ~~~~~~~
        - 2\,\mathrm{Re}\bigl\{ u_k \bar{\mathbf{h}}_k^{\mathsf H} \mathbf{w}_k \bigr\} + 1,
    \end{alignat}
\end{subequations}
where $\tilde{\sigma}_k^2 = \sum\limits_{p=1}^L \sigma_p^2 \left| \bar{\mathbf{h}}_{p,k}^\mathsf{H} \right|^2 + \sigma_{k}^2$ and $\tilde{\sigma}_S^2 = \sum\limits_{p=1}^L \sigma_p^2 \left| \bar{\mathbf{h}}_{p,S}^\mathsf{H} \right|^2 + \sum\limits_{p=1}^L \frac{\sigma_p^2 \left\| \mathbf{H}_{p}^\mathsf{H} \right\|_\text{F}^2}{|\rho_S|^2 \left\| \bar{\mathbf{h}}_{S} \right\|^2} +  \frac{\bar{\sigma}_S^2}{|\rho_S|^2 \left\| \bar{\mathbf{h}}_{S} \right\|^2}$.
Similar to the previous section, we iteratively optimize the beamforming vectors at the BS and amplitude and phase-shift coefficients of STAR-RIS.

\subsection{Solution for Sensing Transmit Beamformers}
To determine the optimal sensing transmit beamformer, while keeping all other variables fixed ($\boldsymbol{\beta}$, $\boldsymbol{\phi}$, $\{ \mathbf{w}_k \}_{k \in \mathcal{K}}$, $\{ u_k \}_{k \in \mathcal{K}}$, $\{ \xi_k \}_{k \in \mathcal{K}}$) except for $\mathbf{w}_S$, we revisit the same optimization problem presented in equation \eqref{eq:opt_tx_eqi_sensing}. In this context, we replace $C_{S,\text{pas}}$ with $C_{S,\text{act}}$, where $C_{S,\text{act}}$ is defined as
\begin{align}
    C_{S, \text{act}} = \max \Big\{&\gamma_{S,\text{req}}\bigg(\sum\limits_{c\in \mathcal{K}} \left| \bar{\mathbf{h}}_{S}^\mathsf{H} \mathbf{w}_{c} \right|^2 +  \sum\limits_{p=1}^L \sigma_p^2 | \bar{\mathbf{h}}_{p,S}^\mathsf{H}|^2  \\[-0.2cm]
    & + \sum\limits_{p=1}^L \frac{\sigma_p^2 \| \mathbf{H}_{p}^\mathsf{H}\|_\text{F}^2}{ |\rho_\text{S}|^2 \| \bar{\mathbf{h}}_{S} \|^2} + \sigma_S^2\bigg), \frac{1}{\gamma_{\text{e},k,\text{req}}}\left| \bar{\mathbf{h}}_{S}^\mathsf{H} \mathbf{w}_{k} \right|^2 \nonumber \\[-0.2cm]
    &  - \hspace{-0.3cm} \sum_{c\in \mathcal{K} \backslash \{k\}} \hspace{-0.3cm} \left| \bar{\mathbf{h}}_{S}^\mathsf{H} \mathbf{w}_{c} \right|^2 - \sum_{p=1}^L \sigma_p^2 \| \bar{\mathbf{h}}_{p,S}^\mathsf{H}\|^2 - \sigma_\text{e}^2 \Big\}.\nonumber
\end{align}

\subsection{Solution for Communication Transmit Beamformers}
With the fixed $\boldsymbol{\beta}$ and $\boldsymbol{\phi}$, Problem $\Problem{3}$ becomes convex with respect to each of the optimization variables $\boldsymbol{\beta}$, $\boldsymbol{\phi}$, and $\{ \mathbf{w}_k \}$. The key differences between the transmit beamformers for passive STAR-RIS and active STAR-RIS are twofold: i) the expressions for the noise variance in the SINRs of communication and sensing; and ii) additional power constraint on active STAR-RIS. 
With fixed $\boldsymbol{\beta}$ and $\boldsymbol{\phi}$, the noise variance terms become constant. The problem then reduces to the previous case, giving the optimal ${u_k}$ and ${\xi_k}$.
To handle the active STAR-RIS power constraint and find the optimal $\mathbf{w}_k$, we define the Lagrangian of Problem $\Problem{3}$ as
\begin{align}
    &L_{\text{act}}( \{ \mathbf{w}_k\}, \lambda, \boldsymbol{\mu}, \boldsymbol{\nu} ) 
    = 
    \tilde{L}_{\text{pas}}( \{ \mathbf{w}_k\}, \lambda, \boldsymbol{\mu} ) \\
    &\hspace{.44cm} - \sum_{l=1}^L  \nu_l \bigg(\sum_{k \in \mathcal{K} \cup \{ S \}} \hspace{-0.2cm}\| \boldsymbol{\Theta}_{\text{T},l} \mathbf{H}_l \mathbf{w}_k \|^2 + \| \boldsymbol{\Theta}_{\text{T},l} \|_\text{F}^2 \sigma_l^2 \nonumber \\[-0.2cm]
    &\hspace{2cm} \hspace{-7pt} + \hspace{-0.35cm}\sum_{k \in \mathcal{K} \cup \{ S \}} \hspace{-0.2cm} \| \boldsymbol{\Theta}_{\text{R},l} \mathbf{H}_l \mathbf{w}_k \|^2 + \| \boldsymbol{\Theta}_{\text{R},l} \|_\text{F}^2 \sigma_l^2 - P_{\text{A}, l, \text{max}}\bigg), \nonumber
\end{align}
where $\tilde{L}_{\text{pas}}( \{ \mathbf{w}_k\}, \lambda, \boldsymbol{\mu} )$ is the function from \eqref{eq:lagrange} with $\sigma_k$ and $\sigma_S$ replaced by $\tilde{\sigma}_k$ and $\tilde{\sigma}_S$, respectively. Obtaining the first-order derivative of the Lagrange function $L_{\text{act}}$, we get the optimal transmit beamformer as
\begin{align}
\label{eq:tx_beamformer_act}
    &\hspace{-0.2cm} \mathbf{w}_{k, \text{act}} = \Big(\hspace{-0.2cm} \sum_{\ell \in \mathcal{K} \cup \{ S \}} \hspace{-0.2cm} \alpha_\ell \xi_\ell |u_\ell|^2 \bar{\mathbf{h}}_\ell \bar{\mathbf{h}}_\ell^\mathsf{H} + \lambda \mathbf{I}
    - \hspace{-0.2cm}\sum_{\ell \in \mathcal{K} \cup \{ S \} } \hspace{-0.2cm}\mu_\ell \bar{\gamma}_{\ell,\text{req}}^{(k)} \bar{\mathbf{h}}_S \bar{\mathbf{h}}_S^\mathsf{H} \nonumber \\[-0.2cm]
    &\hspace{-0.2cm} - \sum_{p=1}^L \nu_p \mathbf{H}_p^\mathsf{H} 
    (\boldsymbol{\Theta}_{\text{T},p}^\mathsf{H} \boldsymbol{\Theta}_{\text{T},p} + \boldsymbol{\Theta}_{\text{R},p}^\mathsf{H} \boldsymbol{\Theta}_{\text{R},p})  \mathbf{H}_p \Big)^{-1} \alpha_k \xi_k u_k^{*} \bar{\mathbf{h}}_k.
\end{align}
Thus, we can construct the following dual problem as
\begin{subequations}%
\label{opt:opt_tx_eqi_act_dual}
    \begin{alignat}{3}
        & && \max_{\lambda,\,\boldsymbol{\mu},\,\boldsymbol{\nu}} && 
        \tilde{L}_{\text{act}}(\lambda,\,\boldsymbol{\mu},\,\boldsymbol{\nu}) 
        = L_{\text{act}}\bigl( \{\mathbf{w}_{k,\text{act}}\},\,\lambda,\,\boldsymbol{\mu},\,\boldsymbol{\nu} \bigr), 
        \\ 
        & && \mathrm{~~s.t.} && 
        \lambda,\,\mu_l,\,\nu_p \ge 0, \; \forall\, l,p.
    \end{alignat}
\end{subequations}
In the case of dual variable $\lambda$, we can employ the same procedure proposed in Section~\ref{subsec:solution_tx_beamformer} to determine the optimal value. In addition, we utilize the sub-gradient descent method to update dual variable $\nu_p$ by
\begin{align}
    \nu_p \leftarrow \nu_p + &\delta \Big(\hspace{-0.1cm} \sum_{k \in \mathcal{K} \cup \{ S \}} \hspace{-0.3cm} \| \boldsymbol{\Theta}_{\text{T},p} \mathbf{H}_p \mathbf{w}_{k, \text{act}} \|^2 + \| \boldsymbol{\Theta}_{\text{T},p} \|_\text{F}^2 \sigma_p^2 
    \\[-0.1cm]
    & \hspace{0.22cm} \hspace{-12pt} + \hspace{-0.3cm} \sum_{k \in \mathcal{K} \cup \{ S \}} \hspace{-0.3cm} \| \boldsymbol{\Theta}_{\text{R},p} \mathbf{H}_p \mathbf{w}_{k, \text{act}} \|^2 + \| \boldsymbol{\Theta}_{\text{R},p} \|_\text{F}^2 \sigma_p^2 - P_{\text{A}, l, \text{max}}\Big), \nonumber
\end{align}
where $\delta$ is the learning rate. After the values of the dual variables ($\lambda$, $\boldsymbol{\mu}$, $\boldsymbol{\nu}$) converge, we can obtain the optimal solution $\mathbf{w}_k^{\text{act, opt}}$ by substituting the values of the dual variables into \eqref{eq:tx_beamformer_act}. 
We remark that the duality gap between \eqref{eq:opt_tx_eqi_act} and \eqref{opt:opt_tx_eqi_act_dual} is zero as Problem $\Problem{3}$ is convex with respect to $\mathbf{w}_k$. This ensures that the obtained solution is globally optimal and satisfies all the constraints of the problem in \eqref{eq:opt_tx_eqi_act}.

\subsection{Solution for Amplitude and Phase Shift Coefficients}
Similarly to the previous section, we fix the group of variables $\{ \{\mathbf{w}_k\}, \{u_k\}, \{ \xi_k \} \}$ and optimize the amplitude and phase shifts of the STAR-RIS. Specifically, we hold the T\&R coefficient matrices for STAR-RIS $1,\ldots, q-1, q+1, \ldots, L$ constant and sequentially optimize the T\&R coefficient matrices of the $q$-th STAR-RIS. The effective channel from STAR-RIS $p$ to user $k$ can be represented as
\begin{align}
    \mathbf{\bar{h}}_{p,k}^\mathsf{H} &=  \bar{\boldsymbol{\theta}}_q^\mathsf{T} \bar{\mathbf{F}}_{p,k,q} + \mathbf{b}_{p,k,q}^\mathsf{H} \nonumber \\
    &= \boldsymbol{\theta}_{\text{T},q}^\mathsf{T} \mathbf{F}_{\text{T},p,k,q} + \boldsymbol{\theta}_{\text{R},q}^\mathsf{T} \mathbf{F}_{\text{R},p,k,q} + \mathbf{b}_{p,k,q}^\mathsf{H},
\end{align}
where $\bar{\mathbf{F}}_{p,k,q} = [\mathbf{F}_{\text{T},p,k,q}^\mathsf{T}, \mathbf{F}_{\text{R},p,k,q}^\mathsf{T}]^\mathsf{T}$, and $\mathbf{F}_{\text{T},p,k,q}$, $\mathbf{F}_{\text{R},p,k,q}$, and $\mathbf{b}_{p,k,q}$ are defined similarly in \ref{subsec:passive_channel_opt}. The detailed derivation is provided in Appendix \ref{appendix:derivation_channel}.

Then, we can formulate the optimization problem as 
\begin{subequations}%
\label{eq:opt_act_ris}
    \begin{alignat}{3}
        & && \min_{\tilde{\boldsymbol{\Theta}}_q} && 
        \mathrm{tr}\bigl( \tilde{\boldsymbol{\Theta}}_q \mathbf{J}_q \bigr) 
        + \sum_{k\in\mathcal{K}} |u_k|^2\, \mathrm{tr}\bigl( \tilde{\boldsymbol{\Theta}}_q \tilde{\mathbf{N}}_{k,q} \bigr), 
        \\ 
        & && \mathrm{~~s.t.~} && 
        \mathrm{tr}\bigl( \tilde{\boldsymbol{\Theta}}_q \mathbf{M}_{k,q} \bigr) 
        \le 
        \mathrm{tr}\bigl( \tilde{\boldsymbol{\Theta}}_q \tilde{\mathbf{N}}_{S,q} \bigr) + \sigma_{\text{e}}^2, 
        \; k\in\mathcal{K}, 
        \label{subeq:act_reform_opt_ris_b} \\ 
        & && && 
        \mathrm{tr}\bigl( \tilde{\boldsymbol{\Theta}}_q \mathbf{M}_{S,q} \bigr) 
        \ge \tilde{\sigma}_S^2, 
        \label{subeq:act_reform_opt_ris_c} \\ 
        & && && 
        \mathrm{tr}\bigl( \tilde{\boldsymbol{\Theta}}_q \boldsymbol{\Xi}_p \bigr) 
        \le P_{\text{A},p}^{\text{max}}, 
        \label{subeq:act_reform_opt_ris_d} \\ 
        & && && 
        \eqref{subeq:reform_opt_ris_e} - \eqref{subeq:reform_opt_ris_g},
    \end{alignat}
\end{subequations}
where $\tilde{\mathbf{N}}_{k,q}$, $\boldsymbol{\Xi}_p$, and $\mathbf{V}_p$ are defined as
\begin{align}
    \tilde{\mathbf{N}}_{k,q} &= \sum_{p=1}^L \sigma_p^2
    \begin{pmatrix}
    \bar{\mathbf{F}}_{p,k,q} \bar{\mathbf{F}}_{p,k,q}^\mathsf{H} & \bar{\mathbf{F}}_{p,k,q} \mathbf{b}_{p,k,q} \\
    \mathbf{b}_{p,k,q}^\mathsf{H} \bar{\mathbf{F}}_{p,k,q}^\mathsf{H} & \mathbf{b}_{p,k,q}^\mathsf{H} \mathbf{b}_{p,k,q}
    \end{pmatrix}, \\
    \boldsymbol{\Xi}_p &=
    \begin{pmatrix}
    \mathbf{V}_p & \mathbf{0} & \mathbf{0} \\
    \mathbf{0} & \mathbf{V}_p & \mathbf{0} \\
    \mathbf{0} & \mathbf{0} & 0
    \end{pmatrix}, \\
    \mathbf{V}_p &= \sum_{k \in \mathcal{K} \cup \{ S \}} \mathrm{diag}(\mathbf{w}_k^\mathsf{H} \mathbf{H}_p^\mathsf{H}) \mathrm{diag}(\mathbf{H}_p \mathbf{w}_k) + \sigma_p^2 \mathbf{I}.
\end{align}
For some matrices $\mathbf{A}$, $\mathbf{B}$, and a diagonal matrix $\boldsymbol{\Theta} = \mathrm{diag} (\boldsymbol{\theta})$, $\| \mathbf{A} \boldsymbol{\Theta} \mathbf{B} \|_\text{F}^2 = \boldsymbol{\theta}^\mathsf{T} (\mathbf{A}^\mathsf{H} \mathbf{A} \circ \mathbf{B} \mathbf{B}^\mathsf{H}) \boldsymbol{\theta}^{*}$ holds.
Then, if $q < p$, $\| \mathbf{H}_p \|_\text{F}^2 $ can be represented as
\begin{align}
    \| \mathbf{H}_p \|_\text{F}^2 = \mathrm{tr} ( \tilde{\boldsymbol{\Theta}}_q \mathbf{Q}_{q,p} ),
\end{align}
where $\mathbf{Q}_{q,p}$ is defined as
\begin{align}
    \mathbf{Q}_{q,p} &= \begin{pmatrix}
   (\tilde{\mathbf{T}}_{q,p}^\mathsf{H} \tilde{\mathbf{T}}_{q,p}) \circ (\mathbf{H}_{q} \mathbf{H}_{q}^\mathsf{H}) & \mathbf{0} & \boldsymbol{\omega}_{q,p} \\
    \mathbf{0} & \mathbf{0} & \mathbf{0} \\
    \boldsymbol{\omega}_{q,p}^\mathsf{H} & \mathbf{0} & \| \mathbf{H}_p - \mathbf{H}_{q,p} \mathbf{H}_{q} \|_\text{F}^2
    \end{pmatrix}, \\
    \boldsymbol{\omega}_{q,p} &= \mathrm{diag} (\mathbf{H}_{q}(\mathbf{H}_p^\mathsf{H} - \mathbf{H}_{q}^\mathsf{H} \mathbf{H}_{q,p}^\mathsf{H} ) \tilde{\mathbf{T}}_{q,p} ).
\end{align}
Then, the term $\tilde{\sigma}_S$ can be represented with respect to $\tilde{\boldsymbol{\Theta}}_q$ as
\begin{align}
   \tilde{\sigma}_S^2 = \mathrm{tr} (\tilde{\boldsymbol{\Theta}}_q \tilde{\mathbf{N}}_{S,q} ) + \frac{ \mathrm{tr}( \tilde{\boldsymbol{\Theta}}_q \tilde{\mathbf{Q}}_q ) + \bar{\sigma}_S^2}{|\rho_S|^2 \mathrm{tr}(\tilde{\boldsymbol{\Theta}}_q \mathbf{N}_q ) },
\end{align}
where $\tilde{\mathbf{Q}}_q = \sum_{p=1}^L \sigma_p^2 \mathbf{Q}_{q,p}$. 

Constraint \eqref{subeq:act_reform_opt_ris_c}, characterized by the fractional programming term, is inherently non-convex. To transform this non-convex constraint into a more tractable form, we first reformulate constraint \eqref{subeq:act_reform_opt_ris_c} as 
\begin{align}
    \hspace{-0.4cm}|\rho_S|^2 \mathrm{tr}(\tilde{\boldsymbol{\Theta}}_q \mathbf{N}_q ) \mathrm{tr}(\tilde{\boldsymbol{\Theta}}_q (\mathbf{M}_{S,q} - \tilde{\mathbf{N}}_{S,q}) ) \geq  \mathrm{tr}( \tilde{\boldsymbol{\Theta}}_q \mathbf{Q}_q ) + \bar{\sigma}_S^2.
\end{align}
It can be observed the left-hand side is still non-convex. We address this via SCA. Specifically, we define $g(\tilde{\boldsymbol{\Theta}}_q) \triangleq |\rho_S|^2 \mathrm{tr}(\tilde{\boldsymbol{\Theta}}_q \mathbf{N}_q ) \mathrm{tr}(\tilde{\boldsymbol{\Theta}}_q (\mathbf{M}_{S,q} - \tilde{\mathbf{N}}_{S,q}) )$ and construct an approximation of $g(\tilde{\boldsymbol{\Theta}}_q)$, which is given as
\begin{align}
    g(\tilde{\boldsymbol{\Theta}}_q) \approx& g(\tilde{\boldsymbol{\Theta}}_q^{(i)}) + \mathrm{tr} \Big[\Big((\mathbf{M}_{S,q} - \tilde{\mathbf{N}}_{S,q}) \mathrm{tr}(\tilde{\boldsymbol{\Theta}}_q^{(i)} \mathbf{N}_q )  \\
    &+ \mathrm{tr}\big(\tilde{\boldsymbol{\Theta}}_q^{(i)} (\mathbf{M}_{S,q} - \tilde{\mathbf{N}}_{S,q}) \big) \mathbf{N}_q \Big)^\mathsf{T} (\tilde{\boldsymbol{\Theta}}_q - \tilde{\boldsymbol{\Theta}}_q^{(i)})\Big], \nonumber
\end{align}
where $\tilde{\boldsymbol{\Theta}}_q^{(i)}$ is the local point obtained at the $i$-th iteration. Consequently, problem \eqref{eq:opt_act_ris} can be approximated as
\begin{subequations}%
\label{eq:act_reform_opt_ris_approx}
    \begin{alignat}{3}
        & && \min_{\tilde{\boldsymbol{\Theta}}_q} ~~ &&
        \mathrm{tr}\bigl( \tilde{\boldsymbol{\Theta}}_q \mathbf{J}_q \bigr) 
        + \mathrm{tr}\bigl( \tilde{\boldsymbol{\Theta}}_q \tilde{\mathbf{N}}_q \bigr), 
        \\ 
        & && \mathrm{~~s.t.~} && 
        g\bigl( \tilde{\boldsymbol{\Theta}}_q \bigr) 
        \ge 
        \mathrm{tr}\bigl( \tilde{\boldsymbol{\Theta}}_q \mathbf{Q}_q \bigr) + \bar{\sigma}_S^2, 
        \\ 
        & && && 
        \eqref{subeq:act_reform_opt_ris_b},~\eqref{subeq:act_reform_opt_ris_d},~\eqref{subeq:reform_opt_ris_e} - \eqref{subeq:reform_opt_ris_g}.
    \end{alignat}
\end{subequations}
The optimal solution to the problem \eqref{eq:act_reform_opt_ris_approx} can be obtained efficiently using standard convex optimization techniques.

\section{Simulation Results}

\begin{table}
    \centering
    \caption{System Parameters}
    \label{table:spec_RIS}
    \resizebox{\if 1\mycmd 0.6 \else 0.95 \fi \columnwidth}{!}{
    \begin{tabular}{ccc}
    \toprule
         \textbf{Parameter} & \textbf{Symbol} & \textbf{Value} \\ \midrule[\heavyrulewidth]\midrule[\heavyrulewidth]
         Number of antennas & $N$ & $20$ \\ 
         Transmit power & $P_\text{BS}$ & $30~\text{dB}$ \\ 
         Number of communication users & $K$ & $6$ \\ 
         Number of RISs & $L$ & 3 \\
         Number of RIS elements & $M$ & 64 \\ 
         Noise power & $\sigma_S^2 = \sigma_k^2 = \sigma_\text{e}^2$ & $-110~\text{dB}$ \\ 
         RCS of the sensing target & $\rho^2$ & $0~\text{dB}$ \\ 
         Rician factors & $F_{\mathbf{G}_l}, F_{\mathbf{T}_{p,l}}, F_{\mathbf{h}_{l,k}}$ & $3~\text{dB}$ \\
         Sensing SINR constraint & $\gamma_{S, \text{req}}$ & $-30~\text{dB}$ \\
         Leakage SINR constraint & $\gamma_{\text{e},k,\text{req}}$ & $5~\text{dB}$ \\
    \bottomrule
    \end{tabular}
    }
\label{table:parameter_symbol_value_AI_radcom}
\vspace{-6pt}
\end{table}

For simulations, the key system parameters are listed in Table \ref{table:spec_RIS}. 
It is worth noting that the proposed method is not limited to the selected parameters. 
To clearly represent the positions of the BS and the STAR-RISs, we adopt a three-dimensional Cartesian coordinate system. The BS is placed at the origin $(0,0,25)~\text{meter} (\text{m})$, whereas the three active STAR-RISs are located at $(10,0,0)$, $(20,0,0)$, and $(30,0,0)~\text{m}$, respectively. Thus, the horizontal interval between STAR-RISs is set to $10~\text{m}$. For each STAR-RIS, two communication users are randomly dropped on its reflection side, resulting in $K = 6$ users in total. The sensing target is assumed to hover at $(32,0,0)~\text{m}$. The RIS channel follows a Rician channel model as described in Appendix \ref{appendix:rician_channel_modeling} in detail. Unless stated otherwise, all simulations use the parameters described above.

\subsection{Impact of Sensing SINR Constraint}

\begin{figure}[t]
    \centering
    \includegraphics[draft=false, width= \if 1\mycmd 0.6 \else 0.9 \fi \columnwidth]{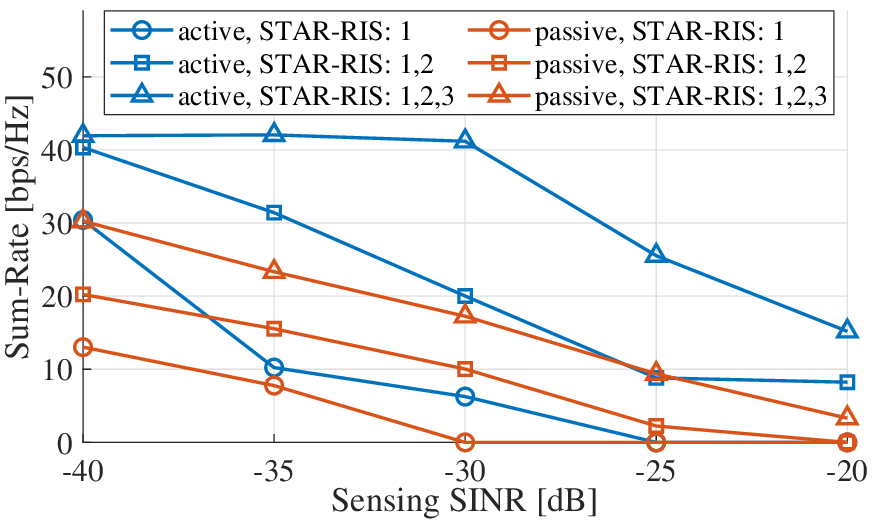}
    \vspace{-8pt}
    \caption{Achievable sum-rate versus required sensing SINR $\gamma_{S,\mathrm{req}}$ for different numbers of STAR-RISs under both active and passive configurations.}
    \label{fig:SR_sensingSINR}
    \vspace{-10pt}
\end{figure}

Fig.~\ref{fig:SR_sensingSINR} illustrates the achievable downlink sum-rate versus the required sensing SINR, $\gamma_{S,\text{req}}$, for both passive and active STAR-RIS deployments comprising one, two, and three cooperative STAR-RISs. 
The label ``passive, STAR-RIS: 1,2'' indicates that STAR-RISs 1 and 2 operate in passive mode, while STAR-RIS 3 remains inactive (i.e., turned off).
To enable a fair comparison, the total transmit power budget is fixed as $P_{\text{total,max}} = P_{\text{BS,max}} + \sum_{l=1}^{L} P_{\text{A},l,\text{max}}$.  
In the passive benchmark cases, the entire power budget is allocated to the BS, i.e., $P_{\text{total,max}} = P_{\text{BS,max}}$. For the active schemes, $40~\%$ of the total power is allocated to the BS ($P_{\text{BS,max}} = 0.4 \times P_{\text{total,max}}$), while the remaining $60~\%$ is equally distributed among the $L$ active STAR-RISs ($P_{\text{A},l,\text{max}} = 0.6 \times P_{\text{total,max}}/L$). Although this power allocation is configured as an exemplary setting, it clearly highlights the relative performance gain of the proposed method over conventional schemes. 

As $\gamma_{\text{S,req}}$ increases, the beamforming optimization at the BS becomes increasingly constrained, requiring more power to be directed toward the radar probing beam. This illustrates the fundamental trade-off between sensing and communication performance.
The simulation result also demonstrates that cooperation among multiple STAR-RISs leads to significant performance improvements. Specifically, increasing the number of active STAR-RISs from one to three yields a $486~\%$ sum-rate gain at $\gamma_{\mathrm{S,req}} = -30\,\mathrm{dB}$. This improvement is primarily attributed to the richer set of cascaded channels provided by the multiple STAR-RISs. Even in the presence of severe fading, blockage, or link outages, data transmission can still be robustly maintained through alternative paths. Furthermore, the cooperative deployment introduces greater spatial path diversity, enabling more robust and flexible joint optimization of beamforming across multiple STAR-RISs and the BS. This diversity mitigates the effects of unfavorable channel conditions and leads to more balanced SINR distributions among users, ultimately improving the overall system throughput.

In addition, active STAR-RISs consistently outperform their passive counterparts due to the presence of low-noise amplifiers at each element. These amplifiers compensate for the inherent multi-hop path-loss of the BS-RIS-user links, thereby restoring or enhancing the transmit power. Additionally, the amplification relaxes the unit-modulus constraint imposed on passive elements, enlarging the feasible beamforming space. As a result, active STAR-RISs enable more flexible and effective beamforming strategies, leading to higher coherent combining gains and ultimately improved sum-rate performance under the same total power constraint.

\subsection{Impact of Information-Leakage Constraint}

\begin{figure}[t]
    \centering
    \includegraphics[draft=false, width= \if 1\mycmd 0.6 \else 0.9 \fi \columnwidth]{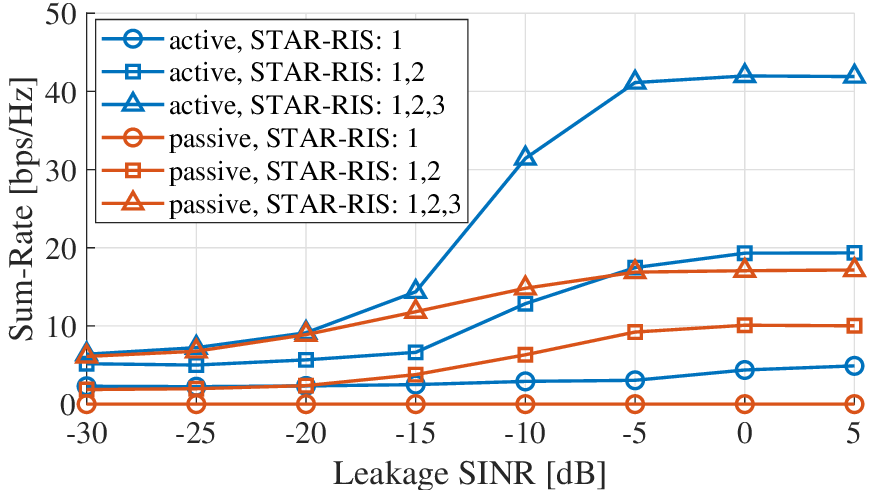}
    \vspace{-8pt}
    \caption{Sum-rate performance versus leakage SINR.}
    \label{fig:SR_leakageSINR}
    \vspace{-10pt}
\end{figure}

Fig.~\ref{fig:SR_leakageSINR} shows the sum-rate versus the maximum tolerable leakage SINR at the sensing target, denoted by $\gamma_{\text{e},k,\mathrm{req}}$. For simplicity, we assume a common leakage SINR constraint, i.e., $\gamma_{\text{e},k,\text{req}}$ is identical for all $k \in \mathcal{K}$.
In the low leakage SINR regime, deploying multiple STAR-RISs proves significantly more advantageous than relying on active STAR-RIS elements. This is because, when strong channel correlation exists between the sensing target and user locations, a single STAR-RIS is more likely to cause elevated leakage SINR due to limited spatial separation in its cascaded links. In contrast, multiple STAR-RISs offer diverse and potentially uncorrelated propagation paths, allowing the optimization algorithm to selectively exploit low-correlation links. This spatial flexibility helps effectively suppress the leakage signal toward the sensing target while maintaining high communication rates.

On the other hand, under a relatively high leakage SINR constraint, the deployment of three active STAR-RISs outperforms other schemes in terms of sum-rate. As the secrecy constraint is relaxed, more transmission energy can be directed toward legitimate users, thereby enhancing their achievable sum-rates.
Furthermore, by utilizing a single radar probing signal, the proposed scheme achieves a low leakage SINR of approximately $-5~\text{dB}$, while maintaining comparable sum-rate performance with negligible or no degradation.


\subsection{Number of STAR-RIS Elements}

\begin{figure}[t]
    \centering
    \includegraphics[draft=false, width= \if 1\mycmd 0.6 \else 0.9 \fi \columnwidth]{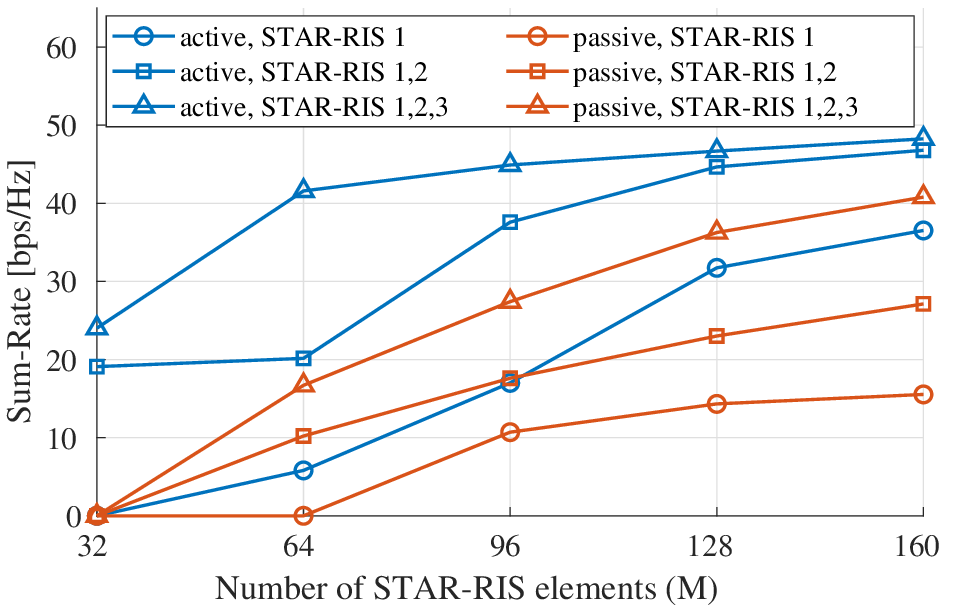}
    \vspace{-10pt}
    \caption{Sum-rate performance versus the number of elements per STAR-RIS.}
    \label{fig:SR_RISElement}
    \vspace{-10pt}
\end{figure}

Fig.~\ref{fig:SR_RISElement} depicts the sum-rate versus the number of reflecting elements per STAR-RIS, denoted by $M$. Across all schemes, a monotonic throughput rise is observed because enlarging not only the DoF in optimizing the beamformer of the STAR-RISs but also array and phase‐shifting gains.
Interestingly, the growth rate under active STAR‑RISs remains noticeably steeper than that of the passive counterpart, although thermal noise rises with the number of RIS elements. This advantage stems from the fact that the per‑element amplification gain, which coherently scales with the array’s beamforming gain exceeds the linearly increasing thermal noise. In addition, the amplification stage loosens the unit-modulus constraint, offering a richer beamforming search space.

\subsection{Interval Between STAR-RISs}
\begin{figure}[t]
    \centering
    \includegraphics[draft=false, width= \if 1\mycmd 0.6 \else 0.9 \fi \columnwidth]{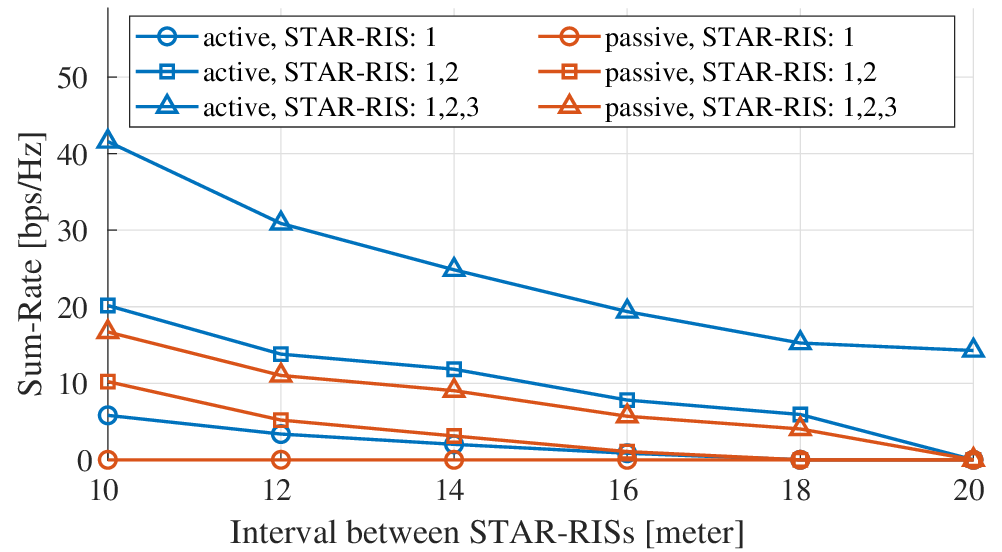}
    \vspace{-8pt}
    \caption{Sum-rate performance versus interval between STAR-RISs.}
    \label{fig:SR_Interval}
    \vspace{-10pt}
\end{figure}

In practical deployments, the spatial separation between multiple STAR-RISs may vary due to constraints in site availability, infrastructure layout, or mobility scenarios. To investigate the impact of such variations on system performance, Fig.~\ref{fig:SR_Interval} presents the sum-rate as a function of the horizontal interval between STAR-RISs under both active and passive configurations. As the interval increases, the sum-rate performance tends to degrade, which can be attributed to the increasing path loss.
The degradation is particularly significant in passive STAR-RIS scenarios, where the absence of signal amplification leads to poor resilience against increased separation. On the other hand, active STAR-RISs demonstrate a relatively robust performance, especially when multiple RISs are deployed. For example, with three active STAR-RISs, the system maintains a high sum-rate even at wider intervals due to the power-assisted signal forwarding that partially compensates for the higher path loss from the BS.

\subsection{Total Transmit Power Budget}

\begin{figure}[t]
    \centering
    \includegraphics[draft=false, width= \if 1\mycmd 0.6 \else 0.9 \fi \columnwidth]{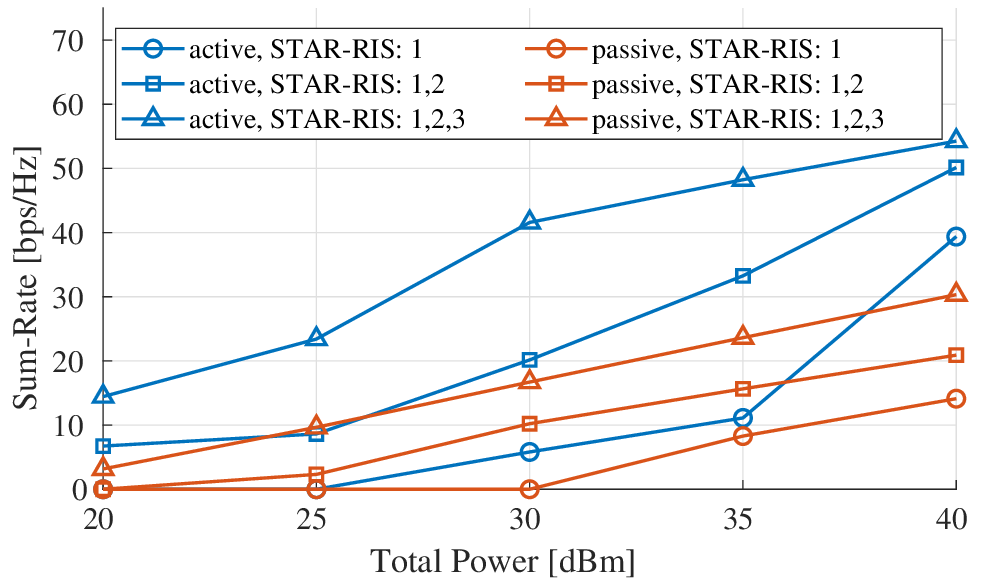}
    \vspace{-8pt}
    \caption{Sum-rate performance versus the total transmit power budget.}
    \label{fig:SR_TotalPower}
    \vspace{-10pt}
\end{figure}

To evaluate the impact of power availability on system performance, Fig.~\ref{fig:SR_TotalPower} presents the sum-rate versus total transmit power under varying numbers of STAR-RISs for both active and passive configurations. As the power increases, all configurations benefit from improved sum-rate, but the active STAR-RISs exhibit significantly greater scalability. In particular, the deployment of multiple active STAR-RISs leads to rapid performance gains due to their ability to amplify signals, effectively utilizing the additional power. Conversely, passive STAR-RISs, lacking amplification, exhibit only modest improvements even at high power levels. These results demonstrate that active STAR-RISs are more power-efficient and better suited for high-throughput scenarios in practical deployments where power budgets are flexible.

\section{conclusion}
This paper proposes the first secure RIS-ISAC system that fuses multiple active, cooperatively operated STAR-RISs, delivering full-space LoS coverage, mitigating multi-hop fading, and hardening physical-layer security. By jointly designing the BS transmit beamformer and the T\&R beamformers of each STAR-RIS through a KKT–guided block-coordinate strategy---combining SCA and SDR---the proposed framework attains high-performance, closed-form solution with modest complexity. Extensive simulations verify that our secure, cooperative, and active STAR-RIS architecture markedly boosts both communication throughput and sensing accuracy while safeguarding confidentiality, all with fewer RIS elements than baselines, revealing its practicality for next-generation ISAC networks.

\begin{appendices}

\section{Rician Channel Modeling}
\label{appendix:rician_channel_modeling}
To model the LoS component in Rician fading between two nodes, we first define the steering vector for angle $\phi$ for representing the array response at the BS as
\begin{align}
    \mathbf{a}(\phi) = [1, \ldots, e^{jkd (N-1) \sin \phi}]^\mathsf{T},
\end{align}
where $k = 2\pi / \lambda$ is the wave number, with $\lambda$ being the signal wavelength, and $d$ is the distance between adjacent elements at the BS.
In addition, let $M_x$ and $M_y$ be the number of horizontal and vertical T\&R elements at each STAR-RIS, respectively, with $M_x M_y = M$. Then, the array response for the UPA at each STAR-RIS for azimuth angle $\phi_a$ and elevation angle $\phi_e$ is expressed as
\begin{align}
    \bar{\mathbf{a}}(\phi_a, \phi_e) = \bar{\mathbf{a}}_x(\phi_a, \phi_e) \otimes \bar{\mathbf{a}}_y(\phi_a, \phi_e),
\end{align}
where $\bar{\mathbf{a}}_x(\phi_a, \phi_e) = [1, \ldots, e^{j k d_x (M_x - 1) \sin\phi_e \cos\phi_a}]^\mathsf{T}$, $\bar{\mathbf{a}}_y(\phi_a, \phi_e) = [
1, \ldots, e^{j k d_y (M_y - 1) \sin\phi_e \sin\phi_a} ]^\mathsf{T}$, and $d_x$ and $d_y$ are the distance between adjacent elements along the x-axis and y-axis, respectively. 
The three channels are defined as
\begin{align}
    \mathbf{G}_l & = \frac{\sqrt{a_{\mathbf{G}_l}}}{d_{\text{BS},l}} \left( \sqrt{\frac{F_{\mathbf{G}_l}}{F_{\mathbf{G}_l} + 1}} \tilde{\mathbf{G}}_{l,\text{L}} + \sqrt{\frac{1}{F_{\mathbf{G}_l} + 1}} \tilde{\mathbf{G}}_{l,\text{N}} \right) , \\ 
    \mathbf{T}_{p,l} & = \frac{\sqrt{a_{\mathbf{T}_{p,l}}}}{d_{p, l}} \left( \sqrt{\frac{F_{\mathbf{T}_{p,l}}}{F_{\mathbf{T}_{p,l}} + 1}} \tilde{\mathbf{T}}_{p,l,\text{L}} + \sqrt{\frac{1}{F_{\mathbf{T}_{p,l}} + 1}} \tilde{\mathbf{T}}_{p,l,\text{N}} \right), \\ 
    \mathbf{h}_{l,k} \hspace{-2pt} & = \hspace{-2pt} \frac{\sqrt{a_{\mathbf{h}_{l,k}}}}{d_{l, k}} \left( \sqrt{\frac{F_{\mathbf{h}_{l,k}}}{F_{\mathbf{h}_{l,k}} + 1}} \tilde{\mathbf{h}}_{l,k,\text{L}} + \sqrt{\frac{1}{F_{\mathbf{h}_{l,k}} + 1}} \tilde{\mathbf{h}}_{l,k,\text{N}} \right), 
\end{align}
where $a_{\mathbf{G}_l}, a_{\mathbf{T}_{p,l}}, a_{\mathbf{h}_{l,k}}$ represent the corresponding free space path-loss excluding the distance terms, $d_{\text{BS},l}, d_{p,l}, d_{l,k}$ are the distance between the BS and STAR-RIS $l$; between STAR-RIS $l$ and $p$ for $p,l=1,\ldots,L$; and between STAR-RIS $l$ and user $k$ in $\mathcal{C}_l$, respectively, $F_{\mathbf{G}_l}, F_{\mathbf{T}_{p,l}}, F_{\mathbf{h}_{l,k}}$ are the corresponding Rician factors, and $\tilde{\mathbf{G}}_{l,\text{L}}, \tilde{\mathbf{T}}_{p,l,\text{L}}, \tilde{\mathbf{h}}_{l,k,\text{L}}$ are defined as
\begin{align}
    \tilde{\mathbf{G}}_{l,\text{L}} &= \bar{\mathbf{a}}(\phi_{\text{BS}, l}^{(A,a)}, \phi_{\text{BS}, l}^{(A,e)}) \mathbf{a}^\mathsf{H} (\phi_{\text{BS},l}^{(D,a)}), \\
    \tilde{\mathbf{T}}_{p,l,\text{L}} &= \bar{\mathbf{a}}(\phi_{p, l}^{(A,a)}, \phi_{p, l}^{(A,e)}) \bar{\mathbf{a}}^\mathsf{H}(\phi_{p, l}^{(D,a)}, \phi_{p, l}^{(D,e)}), \\
    \tilde{\mathbf{h}}_{l,k,\text{L}} &= \bar{\mathbf{a}}^{*}(\phi_{l, k}^{(D,a)}, \phi_{l, k}^{(D,e)}), 
\end{align}
where $\phi_{i,j}^{(A,m)}$ and $\phi_{i,j}^{(D,m)}$ represent the angle of arrival (AoA) and angle of departure (AoD) from node $i$ to node $j$ along the azimuth axis $(m=a)$ or elevation axis $(m=e)$, respectively. $\tilde{\mathbf{G}}_{l,\text{N}}, \tilde{\mathbf{T}}_{p,l,\text{N}}, \tilde{\mathbf{h}}_{l,k,\text{N}}$ denote the NLoS component, whose elements follow $\mathcal{CN}(0,1)$.

\section{Proof for Lemma $1$}
\label{proof:lemma_1}
We define $e_k = \mathbb{E}[ |c_k - \hat{c}_k|^2 ]$, where $\hat{c}_k$ is the estimated symbol at the receiver at user $k$ $(k \in \mathcal{K})$ or the sensing target $(k = S)$. With the receive equalizer $u_k$, the estimated symbol is given by $\hat{c}_k = u_k y_k$. Substituting \eqref{eq:received_signal_comm_k} into the MSE expression of the received communication symbol, we can get
\begin{align}
    e_k = \mathbb{E} \big[ &|c_k - u_k \bar{\mathbf{h}}_k^\mathsf{H}\mathbf{w}_k c_k \nonumber \\&- u_k \Big(\hspace{-0.1cm}\sum_{\ell \in \mathcal{K} \backslash \{ k \}}\hspace{-0.3cm} \bar{\mathbf{h}}_k^\mathsf{H}\mathbf{w}_\ell c_\ell + \bar{\mathbf{h}}_k^\mathsf{H} \mathbf{w}_S s + z_k\Big) |^2 \big].
\end{align}
Then, we can easily derive $e_k$ in \eqref{subeq:opt_eqi_c}. In addition, fixing all the transmit beamformers $\{\mathbf{w}_k\}$ and minimizing $e_k$ with respect to $u_k$ gives the receive equalizer as
\begin{align}
\label{eq:receive_equalizer}
    u_k = \frac{\bar{\mathbf{h}}_{k}^\mathsf{T} \mathbf{w}_{k}^{*}}{\sum_{\ell\in \mathcal{K} \cup \{S\}} \left| \bar{\mathbf{h}}_{k}^\mathsf{T} \mathbf{w}_{\ell}^{*} \right|^2 + \sigma_k^2}.
\end{align}
Plugging \eqref{eq:receive_equalizer} into $e_k$, we can obtain $e_k = \frac{1}{1+\gamma_k}$. Thus, we can alternatively optimize the beamforming vectors $\{ 
\mathbf{w}_k \}$ by minimizing $e_k$ with $\min \log e_k = \max \log (1+\gamma_k)$.

By introducing an auxiliary variable $\xi_k$, the logarithm function in the objective function can be reformulated by
\begin{align}
    \max_{\{\mathbf{w}\}} \log (1+\gamma_k) &= \min_{\{\mathbf{w}\}, \{ u \} } \log e_k \nonumber \\ 
    &= \min_{\substack{\{\mathbf{w}\}, \{ u \}, \{ \xi \} \\ \xi_k > 0}} (\xi_k e_k - \log \xi_k -1).
\end{align}
Ignoring the constant term, we can derive the reformulated optimization problem in \eqref{eq:opt_tx_eqi}.

\section{Proof for Proposition $1$}
\label{proof:prop_1}
We aim to show $\bar{\Delta}(\epsilon) = \frac{\bar{\mathbf{h}}_S^\mathsf{H} (\mathbf{R}_{\text{int}} + \epsilon \mathbf{I})^{-2} \bar{\mathbf{h}}_S}{(\bar{\mathbf{h}}_S^\mathsf{H} (\mathbf{R}_{\text{int}} + \epsilon \mathbf{I})^{-1} \bar{\mathbf{h}}_S)^2}$ does not increase as the Lagrange multiplier $\epsilon$ increases. In what follows, we show that $\frac{d}{d \epsilon} \bar{\Delta}(\epsilon) <0$ for $\epsilon \ge 0$, implying $\bar{\Delta}(\epsilon)$ is non-increasing with respect to $\epsilon$.

We first derive the first-order derivative of $\bar{\Delta}(\epsilon)$ as
\begin{align}
    \frac{d}{d\epsilon} \bar{\Delta}(\epsilon) &= \frac{d}{d\epsilon}(\frac{\bar{\mathbf{h}}_S^\mathsf{H} (\mathbf{R}_{\text{int}} + \epsilon \mathbf{I})^{-2} \bar{\mathbf{h}}_S}{(\bar{\mathbf{h}}_S^\mathsf{H} (\mathbf{R}_{\text{int}} + \epsilon \mathbf{I})^{-1} \bar{\mathbf{h}}_S)^2}) \nonumber \\
    &= \frac{2}{(\bar{\mathbf{h}}_S^\mathsf{H} (\mathbf{R}_{\text{int}} + \epsilon \mathbf{I})^{-1} \bar{\mathbf{h}}_S)^3} f (\epsilon),
\end{align}
where $f(\epsilon)$ is defined as
\begin{align}
    f (\epsilon) &= ( \bar{\mathbf{h}}_S^\mathsf{H} (\mathbf{R}_{\text{int}} + \epsilon \mathbf{I})^{-2} \bar{\mathbf{h}}_S)^2 \nonumber \\ 
    &- (\bar{\mathbf{h}}_S^\mathsf{H} (\mathbf{R}_{\text{int}} + \epsilon \mathbf{I})^{-3} \bar{\mathbf{h}}_S) (\bar{\mathbf{h}}_S^\mathsf{H} (\mathbf{R}_{\text{int}} + \epsilon \mathbf{I})^{-1} \bar{\mathbf{h}}_S).
\end{align}
Since $\mathbf{R}_{\text{int}} + \epsilon \mathbf{I}$ is a positive definite matrix, the value of $\frac{2}{(\bar{\mathbf{h}}_S^\mathsf{H} (\mathbf{R}_{\text{int}} + \epsilon \mathbf{I})^{-1} \bar{\mathbf{h}}_S)^3}$ is always positive. Then the sign of $\frac{d}{d\epsilon} \Delta(\epsilon)$ is determined by $f(\epsilon)$. To prove that $f(\epsilon) \leq 0$, we use a form of the matrix Cauchy-Schwarz inequality, which is given by
\begin{align}
    (\bar{\mathbf{h}}_S^\mathsf{H} (\mathbf{R}_{\text{int}} + \epsilon \mathbf{I})^{-3} \bar{\mathbf{h}}_S) (\bar{\mathbf{h}}_S^\mathsf{H} &(\mathbf{R}_{\text{int}} + \epsilon \mathbf{I})^{-1} \bar{\mathbf{h}}_S) \nonumber \\ 
    &\ge ( \bar{\mathbf{h}}_S^\mathsf{H} (\mathbf{R}_{\text{int}} + \epsilon \mathbf{I})^{-2} \bar{\mathbf{h}}_S)^2.
\end{align}
Thus, we can conclude that $f(\epsilon) \leq 0$, which implies that $\Delta(\epsilon)$ is monotonically non-increasing with respect to $\epsilon$. 

In addition, as $\epsilon$ diverges to $\infty$, we can obtain
\begin{align}
    \Delta(\epsilon) \approx \frac{\bar{\mathbf{h}}_S^\mathsf{H} (\epsilon \mathbf{I})^{-2} \bar{\mathbf{h}}_S}{(\bar{\mathbf{h}}_S^\mathsf{H} (\epsilon \mathbf{I})^{-1} \bar{\mathbf{h}}_S)^2} = \frac{1}{\bar{\mathbf{h}}_S^\mathsf{H} \bar{\mathbf{h}}_S}.
\end{align}
Thus, the function $\Delta(\epsilon)$ has the lower bound of $\frac{1}{\bar{\mathbf{h}}_S^\mathsf{H} \bar{\mathbf{h}}_S}$.

\section{Proof of Proposition $2$}
\label{proof:prop_2}
Since the condition $\gamma_{\text{e},k} \leq \gamma_{\text{e},k,\text{req}}$ with $\mathbf{w}_k (\lambda, \bm{\mu})$ and $\mu_k=0$ is not satisfied, one can readily infer that $f(0) = \frac{1}{\gamma_{\text{e},k,\text{req}}} |\bar{\alpha}_k|^2 |g_{k,k}|^2 - \sum_{\ell \neq k } |\bar{\alpha}_\ell|^2 |g_{k,\ell}|^2 > \sigma_\text{e}^2$. Additionally, the function $f(\mu_k)$ is continuous and differentiable in $(0, 1/(\gamma_{\text{e},k,\text{req}} \kappa_{k,l})]$. Since $f(1/(\gamma_{\text{e},k,\text{req}} \kappa_{k,l})) = -\infty$ due to the denominator terms of the function $f$, the intermediate value theorem implies that there exists at least one solution $\mu_k^{\text{opt}}$ in the range of $(0, 1/(\gamma_{\text{e},k,\text{req}} \kappa_{k,l})]$ such that the equation \eqref{eq:final_mu} is satisfied.
Additionally, the first derivative of $f(\cdot)$ is
\begin{align}
    &f^{'}(\mu_k) = -\frac{2 \kappa_{k,k}}{\gamma_{\text{e},k,\text{req}}} |\bar{\alpha}_k|^2 |g_{k,k}|^2 |\frac{1}{1+\mu_k \kappa_{k,k} }|^3 \\ 
    &- 2 \gamma_{\text{e},k,\text{req}} \kappa_{k,\ell} \sum_{\ell \neq k } |\bar{\alpha}_\ell|^2 |g_{k,\ell}|^2 |\frac{1}{1-\mu_k \gamma_{\text{e},k,\text{req}} \kappa_{k,\ell} }|^3. \nonumber
\end{align}
As $f^{'}(\mu_k) < 0$ in $(0, 1/(\gamma_{\text{e},k,\text{req}} \kappa_{k,l})]$, there exists only one unique solution in that range.

\section{Derivation of $\mathbf{F}_{\text{T},p,k,q}$, $\mathbf{F}_{\text{R},p,k,q}$, $\mathbf{b}_{p,k,q}$}
\label{appendix:derivation_channel}
We first consider the case where $q < p$. $\mathbf{F}_{\text{T},p,k,q}$, $\mathbf{F}_{\text{R},p,k,q}$, and $\mathbf{b}_{p,k,q}$ can be formulated as
\begin{align}
    \mathbf{F}_{\text{T},p,k,q} &= \mathbf{F}_{\text{R},p,k,q} = \mathbf{0}, \\
    \mathbf{b}_{p,k,q} &= \sum_{l=p}^{\hat{l}_k-1} \mathbf{h}_{l,k}^\mathsf{H} \boldsymbol{\Theta}_{\text{T},l} \mathbf{H}_{p,l} + \hspace{-0.3cm} \sum_{l= \max \{ \hat{l}_k, p\}}^{L} \hspace{-0.5cm} \mathbf{h}_{l,k}^\mathsf{H} \boldsymbol{\Theta}_{\text{R},l} \mathbf{H}_{p,l}.
\end{align}
Next, consider the cases of $q \geq p$. With the fixed T\&R coefficient matrices for STAR-RIS $1,\ldots, q-1, q+1, \ldots, L$, we can represent the channel from the STAR-RIS $p$ to STAR-RIS $l$ ($l\ge q $), with respect to $\boldsymbol{\Theta}_{\text{T}, q}$ as
\begin{align}
    \mathbf{H}_{p,l} =
    \tilde{\mathbf{T}}_{q,l} \boldsymbol{\Theta}_{\text{T}, q} \mathbf{H}_{p,q} + (\mathbf{H}_{p,l} - \mathbf{H}_{q,l} \mathbf{H}_{p,q}). 
\end{align}
Then, $\mathbf{F}_{\text{T},p,k,q}$, $\mathbf{F}_{\text{R},p,k,q}$, and $\mathbf{b}_{p,k,q}$ are defined as
\begin{align}
    \mathbf{F}_{\text{T},p,k,q} =& \sum\limits_{l=q}^{\hat{l}_k-1} \mathrm{diag} (\mathbf{h}_{l,k}^{\mathsf{H}} \boldsymbol{\Theta}_{\text{T},l} \tilde{\mathbf{T}}_{q,l} ) \mathbf{H}_{p,q}  \nonumber \\[-.2cm]
    &+ \hspace{-0.6cm}\sum\limits_{l=\max \{ \hat{l}_k,q+1 \} }^L\hspace{-0.7cm} \mathrm{diag} (\mathbf{h}_{l,k}^{\mathsf{H}} \boldsymbol{\Theta}_{\text{R},l} \tilde{\mathbf{T}}_{q,l} ) \mathbf{H}_{p,q},
\end{align}
\vspace{-0.3cm}
\begin{align}
    \mathbf{F}_{\text{R},p,k,q} &= \begin{cases}
    \mathrm{diag}( \mathbf{h}_{q,k}^{\mathsf{H}} ) \mathbf{H}_{p,q}, ~\text{if} ~ q \geq \hat{l}_k, \\
    \mathbf{0}, \hspace{65pt} \text{otherwise}, \end{cases}
\end{align}
\vspace{-0.3cm}
\begin{align}
    & \mathbf{b}_{p,k,q}^\mathsf{H} = \hspace{-0.7cm} \sum_{l=p}^{\min \{ q-1, \hat{l}_k-1 \}}\hspace{-0.7cm} \mathbf{h}_{l,k}^\mathsf{H} \boldsymbol{\Theta}_{\text{T},l} \mathbf{H}_{p,l} 
    +\hspace{-0.2cm} \sum_{l=q+1}^{\hat{l}_k-1} \hspace{-0.1cm} \mathbf{h}_{l,k}^{\mathsf{H}} \boldsymbol{\Theta}_{\text{T},l} (\mathbf{H}_{p,l} - \mathbf{H}_{q,l} \mathbf{H}_{p,q})
    \nonumber \\[-0.1cm]
    &
    + \hspace{-0.6cm}  \sum_{l=\max \{\hat{l}_k,p \}}^{q-1} \hspace{-0.55cm} \mathbf{h}_{l,k}^\mathsf{H} \boldsymbol{\Theta}_{\text{R},l} \mathbf{H}_{p,l} 
    + \hspace{-0.7cm} \sum_{l=\max \{ \hat{l}_k,q+1 \} }^L \hspace{-0.7cm}\mathbf{h}_{l,k}^{\mathsf{H}} \boldsymbol{\Theta}_{\text{R},l} (\mathbf{H}_{p,l} - \mathbf{H}_{q,l} \mathbf{H}_{p,q})
    .
\end{align}

\end{appendices}

\bibliographystyle{IEEEtran}
\bibliography{{IEEEabrv,bibtex}}

\end{document}